\PassOptionsToPackage{table,dvipsnames,svgnames}{xcolor}
\documentclass[11pt,letterpaper]{mystyle}
\usepackage{tcolorbox}
\tcbuselibrary{most,skins,breakable,listings}
\usepackage{fvextra}
\usepackage[comma,authoryear,compress]{natbib}
\usepackage{algorithm}
\usepackage{algorithmicx}
\usepackage{microtype}
\expandafter\def\csname ver@subfig.sty\endcsname{}
\usepackage{float}
\usepackage{bigstrut}
\usepackage{amssymb}
\usepackage{mathtools}
\usepackage{amsthm}
\usepackage{mathrsfs}
\usepackage{nicefrac}
\usepackage{dsfont}
\usepackage{enumitem}
\usepackage{subcaption}
\usepackage{bxcoloremoji}
\usepackage{float}
\usepackage[utf8]{inputenc} %
\usepackage[T1]{fontenc}    %
\usepackage{url}            %
\usepackage{nicefrac}       %
\usepackage{microtype}      %
\usepackage{subcaption}
\usepackage{fdsymbol}
\usepackage{wrapfig}
\usepackage{lipsum}
\usepackage{stackengine}
\usepackage[font=small,labelfont=bf]{caption}
\usepackage{adjustbox}
\usepackage{rotating}
\usepackage{makecell}
\usepackage{threeparttable}
\usepackage[noend]{algpseudocode}
\usepackage{graphicx}
\usepackage{svg}
\usepackage{booktabs}
\usepackage{pifont}
\usepackage{amsfonts}
\usepackage{textcomp}
\usepackage{multirow}
\usepackage{tabularx}
\usepackage{amsmath}
\usepackage{fancyvrb}
\usepackage{alltt}
\usepackage[justification=centering]{caption}
\usepackage{wrapfig}

\newtcolorbox{promptbox}[2][]{ 
    colback=gray!10,
    coltext=black,
    colframe=black,
    fontupper=\normalfont,
    fonttitle=\normalfont, 
    fonttitle=\bfseries\color{black},
    title=#2,                      
    enhanced,
    boxsep=0.2mm,  
    top=1mm,      
    bottom=1mm,   
    left=1mm,
    right=1mm,
    attach boxed title to top left={yshift=-2mm,xshift=2mm},
    boxed title style={
        colback=white,
        colframe=black
    },
    label={#1}
}

\newcommand{\x}{\mathbf{x}}

\definecolor{blanchedalmond}{rgb}{1.0, 0.92, 0.8}
\definecolor{carmine}{rgb}{0.59, 0.0, 0.09}
\definecolor{lightblue}{rgb}{0.22,0.45,0.70}%

\renewcommand{\mathbf}{\boldsymbol}

\makeatletter
\def\Ddots{\mathinner{\mkern1mu\raise\p@
\vbox{\kern7\p@\hbox{.}}\mkern2mu
\raise4\p@\hbox{.}\mkern2mu\raise7\p@\hbox{.}\mkern1mu}}
\makeatother

\definecolor{amaranth}{rgb}{0.9, 0.17, 0.31}
\definecolor{antiquebrass}{rgb}{0.8, 0.58, 0.46}
\definecolor{antiquefuchsia}{rgb}{0.57, 0.36, 0.51}
\definecolor{chromeyellow}{rgb}{0.31, 0.47, 0.26}

\newcommand{\github}{\raisebox{-1.5pt}{\includegraphics[height=1.05em]{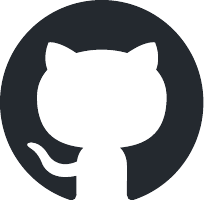}}}

\title{From SFT to RL: Demystifying the Post-Training Pipeline for LLM-based Vulnerability Detection}

\runningtitle{From SFT to RL: Demystifying the Post-Training Pipeline for LLM-based Vulnerability Detection}

\author{
  Youpeng Li$^1$,
  Fuxun Yu$^2$,
  Xinda Wang}

\affil[1]{University of Texas at Dallas}
\affil[2]{Microsoft}

\begin{document}

\begin{abstract}
The integration of large language models (LLMs) into vulnerability detection (VD) has shifted the field toward more interpretable and context-aware analysis. While post-training techniques have shown promise in general coding tasks, their systematic application to VD remains underexplored. In this paper, we present the first comprehensive investigation into the post-training pipeline for LLM-based VD, demonstrating that on-policy reinforcement learning (RL) with GRPO consistently outperforms supervised fine-tuning (SFT), off-policy preference optimization methods, and specialized VD LLMs. Our study further reveals VD-specific post-training guidelines and insights beyond common practices: (1) For data curation, contrary to the widespread use of rationalization-based supervision in prior VD work, SFT based on rejection sampling proves more effective, as rationalization can introduce hallucinations; in RL training, the inherently skewed difficulty distribution of vulnerabilities leads difficulty-aware data filtering to drastically reduce data coverage, causing non-negligible performance loss, and undermines curriculum learning, while pair-based data scheduling can partially mitigate this. (2) For stage interactions, unlike preference optimization typically applied to lightly trained SFT models, increasing SFT epochs consistently benefits off-policy preference optimization in VD tasks; however, excessive SFT suppresses self-exploration in on-policy RL, limiting its gains. (3) For reward mechanisms, naively treating vulnerability classification correctness as reward signals, as commonly adopted in prior work, leads to reward hacking, whereas fine-grained root-cause judgments provide more reliable credit assignment; specification-based rewards further improve efficiency at the cost of additional design and generation effort. (4) For evaluation protocols, traditional binary matching overestimates performance, while LLM-as-a-Judge based on root-cause analysis offers a more robust alternative, albeit with variability across judge models.
\vspace{1cm}

\github{} \textbf{Code Repository}: \href{https://github.com/youpengl/OpenVul}{https://github.com/youpengl/OpenVul}

\coloremojicode{1F917} \textbf{Model Checkpoints \& Datasets}: \href{https://huggingface.co/collections/Leopo1d/openvul}{https://huggingface.co/collections/Leopo1d/openvul}

\coloremojicode{1F4E7} \textbf{Contact}: \href{mailto:youpeng.li@utdallas.edu}{youpeng.li@utdallas.edu}
\end{abstract}

\maketitle

\section{Introduction}
Driven by strong code comprehension capabilities, Large Language Models (LLMs) have been increasingly integrated into vulnerability detection (VD)~\citep{1,2,3,4,5,6,7,8,9,10,11,12,13,14,15,16,17,18,19,20,21,22,23,24,25,26,27,28,29,30,31}. Complementing human experts~\citep{74}, traditional static/dynamic analysis tools~\citep{61,62}, or classic deep learning techniques~\citep{70,71,72}, LLMs offer detailed, interpretable vulnerability reasoning and enable efficient analysis without requiring fully compilable source code. Prompt Engineering was initially introduced to LLM-based VD, evolving from basic zero-shot/few-shot prompting to approaches that incorporate structural information~\citep{1,3,4} or human-curated Chain-of-Thought (CoT) instructions~\citep{2,6,7,21} to enhance vulnerability reasoning. However, the efficacy of these methods is fundamentally bounded by the base models' code understanding and instruction-following capabilities, especially when they lack pre-training on vulnerability-specific tasks~\citep{16,17,18,19,20,21,22,23,24,25,26,27,28,29,30,31}. Furthermore, the defensive concealment of reasoning chains in proprietary models~\citep{74,75}, coupled with the trillion-parameter scales of latest LLMs~\citep{77,78}, restrict their deployment in resource-constrained settings or for sensitive codebase analysis.

Supervised fine-tuning (SFT) has been employed in subsequent works to develop specialized VD LLMs~\citep{53,54,55}. By fine-tuning lightweight models on reasoning datasets comprising CoTs distilled from stronger teacher models, student models have been enabled to mimic reasoning patterns, thereby enhancing VD performance. Nevertheless, the inherent nature of SFT, which trains exclusively on ground-truth answers, limits the model's ability to distinguish between correct and incorrect responses, particularly when dealing with subtle changes before and after vulnerability patching.

Reinforcement Learning from Human Feedback (RLHF) has more recently yielded significant breakthroughs in software engineering tasks~\citep{79}. To mitigate the heavy training overhead and limited generalization of external reward models, Direct Preference Optimization (DPO)~\citep{80} mathematically encodes implicit rewards within its optimization objective, while Odds Ratio Preference Optimization (ORPO)~\citep{81} further eliminates the need for reference models by combining SFT with preference optimization. While these off-policy methods have been applied to VD~\citep{57,58}, 
their reliance on static preference datasets constrains the model's self-exploration capabilities. On-policy methods like Proximal Policy Optimization (PPO)~\citep{82} demonstrate potential to address this, and Group Relative Policy Optimization (GRPO)~\citep{83} further streamlines the process by removing the value model, enhancing both efficiency and generalization.

Despite the proven success of these post-training methods in general domains, their application to VD remains underexplored, with several critical gaps: \textit{\ding{182} Evaluation Scope:} Existing VD empirical studies primarily focus on comparing model performance under limited learning paradigms, whereas the systemic efficacy of various post-training methods remains overlooked. \textit{\ding{183} Data Curation:} The scarcity of high-quality vulnerability reasoning data, the diversity of vulnerability types, and the variance in data difficulty add layers of complexity to training data curation; their collective impact on different post-training stages remains an unsolved puzzle. \textit{\ding{184} Stage Interaction:} Selecting the best-performing checkpoint within a single training stage has become common practice. However, it remains unclear whether the best SFT model from the cold-start stage is also the optimal initialization for subsequent off-policy preference optimization and on-policy RL. \textit{\ding{185} Reward Design:} Coarse-grained binary outcome rewards fail to capture the quality of complex vulnerability reasoning, leaving the impact of reward granularity on RL training efficiency unclear. \textit{\ding{186} Evaluation Protocol:} Traditional binary label comparisons introduce evaluation bias; while LLM-as-a-Judge enables fine-grained assessment, its inherent biases have yet to be systematically measured.

To bridge these gaps, this paper conducts an in-depth examination into the post-training pipeline for LLM-based VD, spanning from cold-start SFT to off-policy preference optimization and on-policy RL, uncovering how data curation, stage interactions, reward systems, and evaluation protocols collectively dictate the efficacy of model training and the accuracy of performance evaluation. Specifically, 
to support a more practical and accurate evaluation, we first curate a context-level VD dataset by enriching high-quality function-level data with inter-procedure contexts extracted via various tools. 
Second, for the cold-start stage, we explore how different SFT data curation methods (rejection sampling vs. rationalization) impact the performance of VD LLMs (\textbf{RQ1}). We then investigate how the intensity of SFT influences subsequent off-policy preference optimization (\textbf{RQ2}) and on-policy RL training (\textbf{RQ3}). 
Third, for the RL stage, we compare full-dataset GRPO training against data filtering based on vulnerability difficulty and introduce two difficulty-aware data scheduling methods, curriculum learning-based and pair-based sampling, to evaluate their impact on performance (\textbf{RQ4}). Additionally, we introduce multi-granularity reward systems, based on detection-level binary comparison, prediction-level CWE matching, reasoning-level root-cause judgment, and sample-level specification alignment, to assess their influence on RL training (\textbf{RQ5}). 
Lastly, regarding evaluation protocols, we quantify the inherent biases of different LLMs acting as LLM-as-a-Judge (\textbf{RQ7}), while demonstrating the discrepancies between binary label-, CWE-, and root cause-based correctness judgment (\textbf{RQ8}).

Through extensive experiments, our study addresses five existing gaps by identifying VD-specific post-training guidelines and insights that go beyond common practices:
\textit{\ding{182}} Models trained using GRPO demonstrate performance that significantly exceeds that of SFT, preference optimization (i.e., DPO, ORPO), and specialized VD LLMs, highlighting the transformative potential of on-policy RL for the post-training of LLMs in VD.
\textit{\ding{183}} In the cold-start stage, SFT with rejection-sampled CoT data is preferred over rationalization, as rationalization introduces hallucinated vulnerability information from teacher models, which misleads learning. In the RL stage, GRPO is highly sensitive to data usage, naive difficulty-based filtering and curriculum learning are less effective due to skewed vulnerability distributions, while pair-based scheduling better supports learning by capturing fine-grained differences between vulnerabilities and patches.
\textit{\ding{184}} LLM-based VD post-training should be treated as a multi-stage pipeline rather than optimizing each stage independently. Use stronger SFT initialization for off-policy preference optimization (DPO/ORPO), but avoid excessive SFT before on-policy RL (GRPO). Over-fitting suppresses exploration and prevents the model from sampling the rollouts it needs for effective learning.
\textit{\ding{185}} Coarse binary rewards mislead RL training and invite reward hacking. Fine-grained signals grounded in root-cause judgments provide reliable credit assignment and yield substantial gains. Specification-based rewards further improve efficiency but require manual rubric design and significant LLM inference costs for generating sample-level specifications.
\textit{\ding{186}} Traditional detection-level binary comparison largely overestimates LLM performance in VD. While prediction-level CWE matching reduces cases of ``correct answer with flawed reasoning,'' it still suffers from false negatives/positives due to CWE's hierarchical structure. Reasoning-level root-cause judgment using LLM-as-a-Judge substantially reduces bias but requires the judge LLM to have strong instruction-following capabilities and domain-specific security expertise.

The contributions of this paper are summarized as follows:

\begin{itemize}[leftmargin=10pt]
\item We are the first to systematically study the post-training pipeline for LLM-based VD from cold-start SFT to off-policy preference optimization and on-policy RL.
\item We examine how data curation, stage interactions, reward mechanisms, and evaluation protocols collectively impact the efficacy of model training and assessment, while identifying key research questions arising from above processes.
\item Based on our evaluation results and findings, we summarize a set of best practices and guidelines to inform and guide future research on LLM-based VD.
\item We open-source a post-training framework \textsc{OpenVul}, including code, datasets, and models, to facilitate future research into the training and evaluation of specialized VD LLMs. 
\end{itemize}
\section{Preliminaries}\label{sec:pre}

\subsection{Problem Formulation.}

\begin{table}[H]{
\small
\setlength{\tabcolsep}{1pt}
\centering
\caption{Objective Functions of LLM Post-training Algorithms}
\label{tab:obj}
\begin{threeparttable}
\begin{tabularx}{\columnwidth}{cl}
\toprule
\textbf{Methods} & \textbf{Objective Functions} $\mathcal{J}(\theta)$ \\ \midrule
\textbf{SFT} & 
\begin{minipage}[t]{0.92\linewidth}
$\begin{aligned}
\mathcal{J}_{SFT}(\theta) = \mathbb{E}_{q, o^+ \sim \mathcal{P(Q, O^+)}} \Bigg[ \frac{1}{|o^+|} \sum_{t=1}^{|o^+|} \log \pi_{\theta}(o^+_t|q, o^+_{<t}) \Bigg]
\end{aligned}$ \refstepcounter{equation}\hfill\textup{(\theequation)}\label{eq:sft}
\end{minipage}
\\ \midrule
\textbf{DPO} & 
\begin{minipage}[t]{0.92\linewidth}
$\begin{aligned} 
\mathcal{J}_{DPO}(\theta) = \mathbb{E}_{\substack{q \sim \mathcal{P(Q)},\\o^+, o^- \sim \pi_{\text{ref}}}_{(\mathcal{O}|q)}} \Bigg[ \log \sigma \bigg( \beta \frac{1}{|o^+|} \sum_{t=1}^{|o^+|} \log \frac{\pi_{\theta}(o_t^+|q, o_{<t}^+)}{\pi_{\text{ref}}(o_t^+|q, o_{<t}^+)} - \beta \frac{1}{|o^-|} \sum_{t=1}^{|o^-|} \log \frac{\pi_{\theta}(o_t^-|q, o_{<t}^-)}{\pi_{\text{ref}}(o_t^-|q, o_{<t}^-)} \bigg) \Bigg] 
\end{aligned}$ \refstepcounter{equation}\hfill\textup{(\theequation)}\label{eq:dpo}
\end{minipage}
\\ \midrule
\textbf{ORPO} & 
\begin{minipage}[t]{0.92\linewidth}
$\begin{aligned} 
\mathcal{J}_{ORPO}(\theta) &= \mathbb{E}_{q, o^+, o^- \sim \mathcal{P(Q, O^+, O^-)}} \Bigg[\frac{1}{|o^+|} \sum_{t=1}^{|o^+|} \log \pi_{\theta}(o_t^+|q, o_{<t}^+) + \beta \log \sigma \left( \log \frac{\text{odds}_{\theta}(o^+|q)}{\text{odds}_{\theta}(o^-|q)} \right) \Bigg]
\end{aligned}$ \refstepcounter{equation}\hfill\textup{(\theequation)}\label{eq:orpo}
\end{minipage}
\\ \midrule
\textbf{\ $\text{GRPO}^*$} & 
\begin{minipage}[t]{0.92\linewidth}
$\begin{aligned} 
\mathcal{J}_{GRPO}(\theta) &= \mathbb{E}_{\substack{q \sim \mathcal{P(Q)},\\\{o_i\}^G_{i=1} \sim \pi_{\text{old}}}_{(\mathcal{O}|q)}} \Bigg[ \frac{1}{G} \sum_{i=1}^G \frac{1}{|o_i|} \sum_{t=1}^{|o_i|}\bigg(\min\bigg[\frac{\pi_{\theta}(o_{i,t}|q_i, o_{i,<t})}{\pi_{\theta_{\text{old}}}(o_{i,t}|q_i, o_{i,<t})} \hat{A}_{i,t}, \text{clip}(\cdot)\bigg] - \beta D_{KL} \left[ \pi_{\theta} || \pi_{\text{ref}} \right] \bigg) \Bigg]
\end{aligned}$ \refstepcounter{equation}\hfill\textup{(\theequation)}\label{eq:grpo}
\end{minipage}
\\ \bottomrule
\end{tabularx}
\begin{tablenotes}[flushleft] 
\item * The $\mathcal{\text{clip}(\cdot)}$ term is negligible in pure on-policy GRPO where $\pi_{\theta} = \pi_{\text{old}}$.
\end{tablenotes}
\end{threeparttable}}
\end{table}

In the context of VD, consider a vulnerability reasoning dataset $\mathcal{D}$ with a data distribution $\mathcal{P(Q, O^+, O^-)}$. Here, $\mathcal{Q}$ represents a set of vulnerability queries, while $\mathcal{O^+}$ and $\mathcal{O^-}$ denote the preferred (chosen) and dispreferred (rejected) responses for each query $q$ in $\mathcal{Q}$, typically generated by prompting strong teacher LLMs.

To enhance the VD capabilities of lightweight student LLMs, a common practice is to perform SFT on a pre-trained model $\pi_\theta$ using training data $\mathcal{P(Q, O^+)}$ composed of queries and chosen responses in Equation~\ref{eq:sft} within Table~\ref{tab:obj}. While SFT enables the model to mimic correct responses, it provides no mechanism for discriminating between valid and flawed outputs, particularly when navigating the subtle nuances of vulnerability patches. To bridge this gap, as shown in Equation~\ref{eq:dpo}, DPO leverages the SFT model as a reference policy $\pi_{\text{ref}}$ and directly optimizes a pairwise preference objective under the Bradley-Terry framework~\citep{80}, eliminating the need for an explicit reward model. ORPO unifies these two stages by combining SFT loss and odds loss, where $\text{odds}_{\theta}(o|q) =\frac{\pi_{\theta}(o|q)}{1 - \pi_{\theta}}(o|q)$ in Equation~\ref{eq:orpo}, further eliminating the need for a reference model to improve training efficiency.

However, both SFT and off-policy preference optimization rely on static supervision. As the model improves, the learned distinctions fail to reflect the latest policy behavior and instead remain derived from outdated training signals, thereby limiting effective self-exploration. To address this, on-policy RL methods have been introduced to incorporate on-the-fly feedback collected from current policy rollouts. In particular, GRPO, as formulated in Equation~\ref{eq:grpo}, samples groups of responses $\{o_i\}^G_{i=1}$ per query $q$ and computes group-normalized relative advantages $\hat{A}_{i,t}$ from their rewards $\{r_i\}^G_{i=1}$ (i.e., $\hat{A}_{i,t} = \tilde{r}_{i}= \frac{r_i - \text{mean}(\{r_i\}^G_{i=1})}{\text{std}(\{r_i\}^G_{i=1})}$), which then guides the on-policy optimization.

\subsection{Context-aware Dataset Construction}

\begin{table}[h]
\caption{Comparison of Representative C/C++ VD Datasets}
\label{tab:vd_datasets}
\centering
\resizebox{\columnwidth}{!}{
\renewcommand{\arraystretch}{1.0}
\begin{tabular}{cccc}
\toprule
\textbf{Open-sourced Dataset} & \textbf{Context Integrity} & \textbf{Data Scale} & \textbf{Label Accuracy} \\ \midrule
\begin{tabular}[c]{@{}c@{}}Devign\citep{95}, BigVul\citep{96}, CVEFixes\citep{97}\\ DiverseVul\citep{72}, MegaVul\citep{48}\end{tabular} & \ding{109} & \ding{108} & \ding{55} \\ \midrule
SVEN\citep{98}, SecLLMHolmes\citep{21} & \ding{109} & \ding{109} & \ding{51} \\ \midrule
PrimeVul\citep{17}, CleanVul\citep{47}, VulRAG\citep{99} & \ding{109} & \ding{108} & \ding{51} \\ \midrule
JitVul\citep{43}, CORRECT\citep{39}, SECODEPLT\citep{100} & \ding{119} & \ding{119} & \ding{51} \\ \midrule
ReposVul\citep{45}, SecVulEval\citep{46} & \ding{119} & \ding{108} & \ding{51} \\ \midrule
OpenVul (Ours) & \ding{108} & \ding{108} & \ding{51} \\ \bottomrule
\end{tabular}
}
\end{table}

As established by recent studies~\citep{32,33,35,37,38,39}, LLMs restricted to isolated functions suffer from semantic blindness. Without broader context such as inter-procedure dependencies and global state behaviors, LLMs often hallucinate assumptions~\citep{32,33,39}, leading to false positives (by overlooking external sanitizers) or false negatives (by missing trigger conditions). However, as shown in Table~\ref{tab:vd_datasets}, most existing VD datasets have the following limitations: (1) function-level VD datasets lack contextual information~\citep{5,47,48,72,17}, which may lead to hallucinated dependencies during LLM reasoning; (2) some datasets, such as JitVul~\citep{43}, ReposVul~\citep{45}, and SecVulEval~\citep{46}, claim to be context-level but lack contextual information for the majority of their samples; (3) several context-level datasets, such as CORRECT~\citep{39}, rely on ground-truth vulnerability locations, such as patch diffs, to guide context extraction, which is inconsistent with real-world scenarios where vulnerability locations are unknown. Other datasets, such as CORRECT~\citep{39} and SECODEPLT~\citep{100}, construct shared contexts for both vulnerable and patched versions, failing to account for newly introduced contexts such as function calls during patching; and (4) certain datasets lack sufficient context-level data to support model post-training~\citep{43,39,100} or even benchmarking, such as SVEN~\citep{98} and SecLLMHolmes~\citep{21}.

To address these issues, we construct a context-aware C/C++ VD dataset, as detailed in Section~\ref{sec:data_split} and Appendix~\ref{sec:datasets}. Building on existing function-level datasets with high labeling accuracy~\citep{17,46}, we exclude irrelevant functions to retain only the target vulnerability-patch pairs. We then apply data deduplication and comment removal to reduce the risk of data leakage. To further enhance data quality and robustness, we adopt a hybrid pipeline that leverages Clang-based~\citep{101} builds for precise dependency resolution, with Joern~\citep{65} serving as a fallback when compilation fails, ensuring broader coverage across diverse codebases. To enrich these function-level code snippets with their surrounding context, we utilize the extracted build information and Code Property Graphs (CPGs) to capture critical dependencies such as global variables, type definitions, macros, and callee functions. This combined strategy enables accurate recovery of code context beyond the local function scope, addressing limitations in prior datasets that lack comprehensive contextual information. Finally, to validate the quality of the constructed dataset, two authors inspect a random subset of 400 samples and observe that the extracted context achieves over 95\% accuracy, demonstrating the reliability of our context construction pipeline.
\section{Post-training Pipeline for LLM-based Vulnerability Detection}\label{sec:pipeline}

\begin{figure}[h]
\centering
\includegraphics[width=0.7\columnwidth]{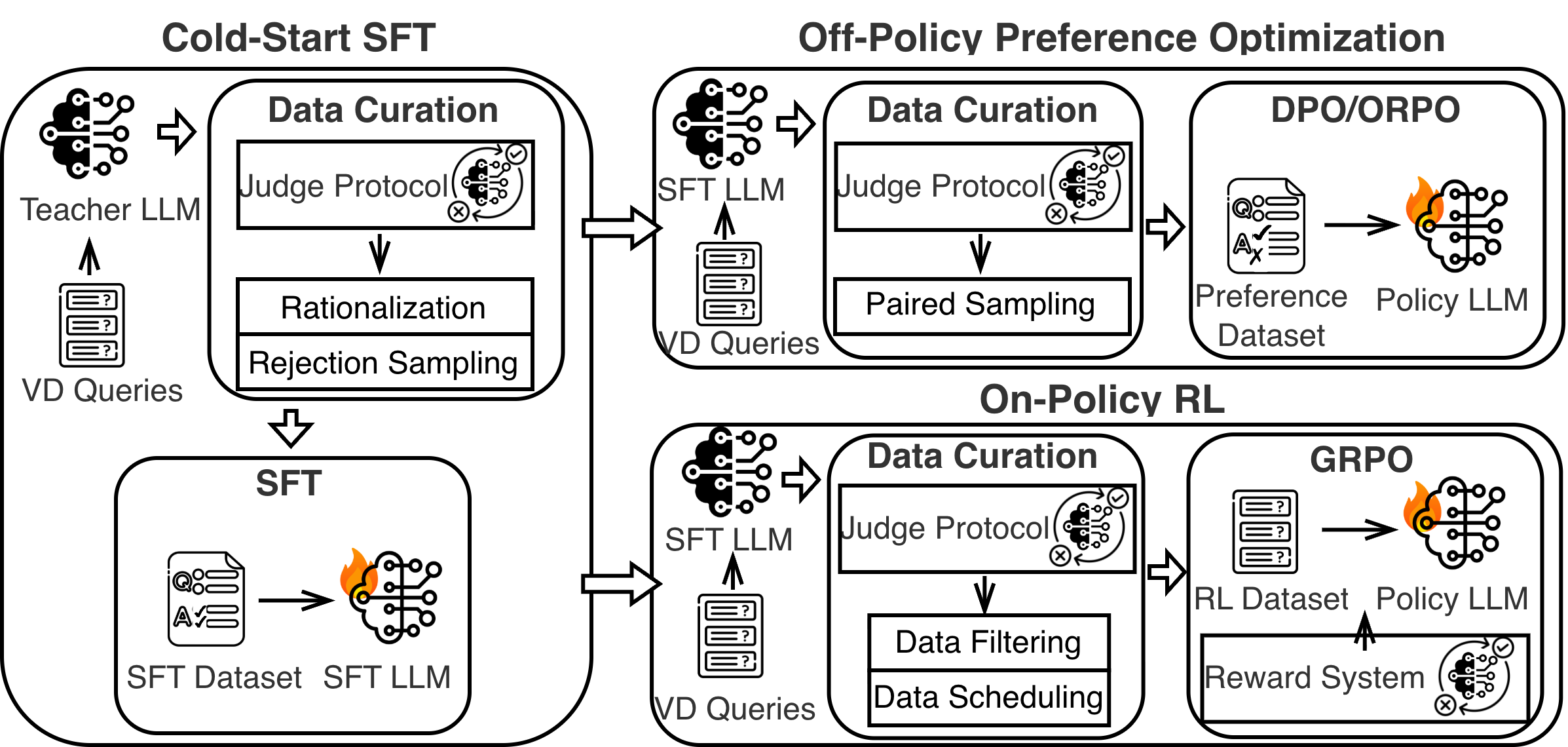}
\caption{Overview of LLM Post-Training Pipelines for VD}
\label{fig:overview}
\end{figure}

Figure~\ref{fig:overview} presents an overview of post-training pipelines for LLM-based VD, including cold-start SFT (Section~\ref{sec:sft}), off-policy preference optimization (Section~\ref{sec:preference_optimization}), and on-policy RL (Section~\ref{sec:rl}). This section examines how data curation, stage interactions, reward design, and evaluation protocols dictate the efficacy of model training and assessment.

\subsection{Cold-start SFT}\label{sec:sft}
\begin{figure}[h]
\centering
\includegraphics[width=0.7\columnwidth]{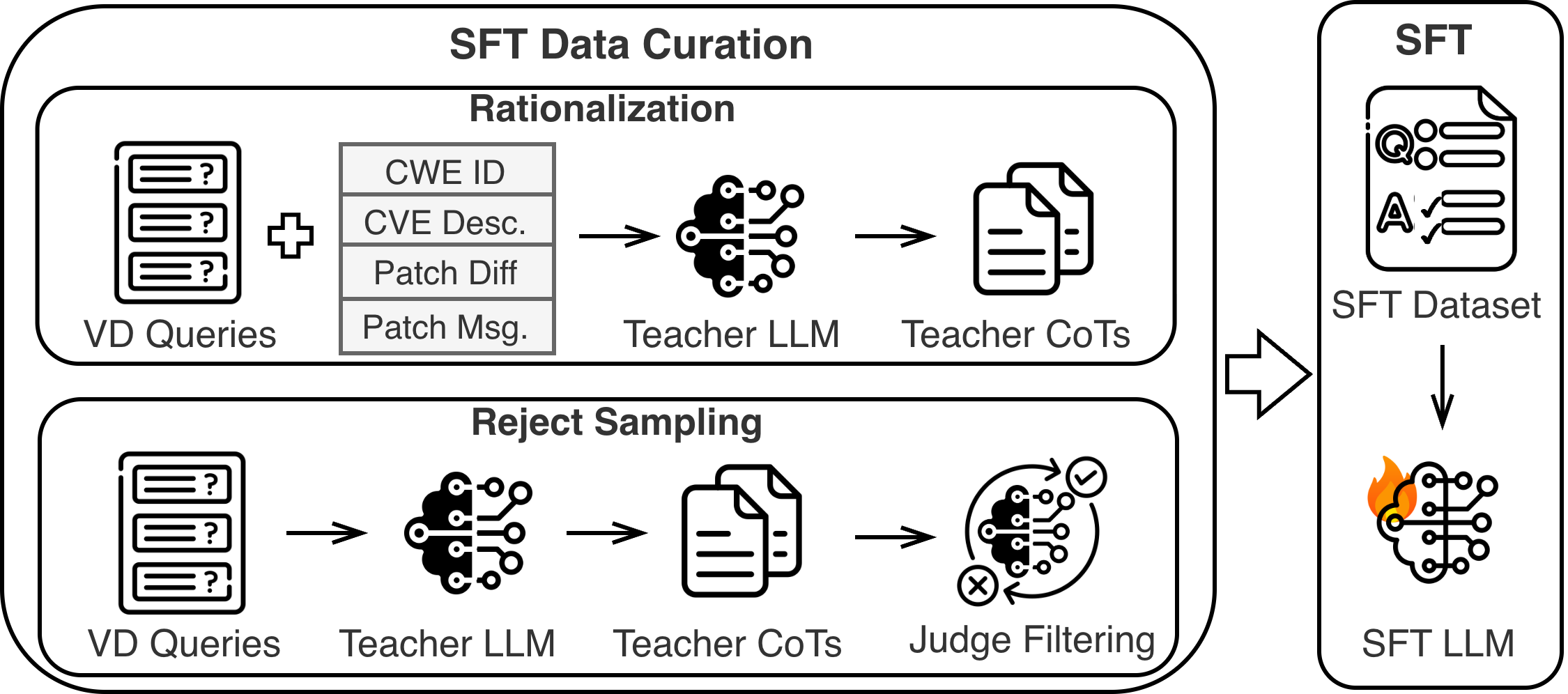}
\caption{Training Pipeline for SFT}
\label{fig:sft}
\end{figure}

\subsubsection{SFT Data Curation}\label{sec:sft_data_curation}
High-quality training data is crucial during the cold-start stage to enhance the model's fundamental VD capabilities. It helps the SFT model better distinguish between vulnerabilities and their corresponding patched versions, thereby improving the effectiveness of subsequent preference optimization and circumventing the sample inefficiency typically triggered by sparse rewards during RL. Figure~\ref{fig:sft} shows two primary methods for curating SFT data:
\begin{itemize}[leftmargin=10pt]
    \item \textit{Rejection Sampling} leverages a dataset of VD queries to prompt a high-capacity teacher LLM to generate multiple candidate CoTs. These candidates are then processed through a rigorous LLM-judge filtering protocol (described in Section~\ref{sec:filter}) to curate an SFT dataset with high-quality vulnerability reasoning CoTs.
    \item \textit{Rationalization} provides teacher LLM with both target code and associated ground-truth metadata, including CWE IDs, vulnerability descriptions, commit messages, and patch diff. By leveraging this gold standard information, the model generates highly accurate vulnerability analyses grounded in factual evidence.
\end{itemize}

These two methods are complementary in their strengths and weaknesses. Rejection sampling uncovers the model's inherent problem-solving logic without external guidance and effectively filters out responses with flawed reasoning. However, it often fails to generate valid reasoning paths for vulnerability samples with subtle or complex characteristics, leading to a high rejection rate where many samples are discarded during the filtering stage. Rationalization ensures a high success rate and strong alignment between the reasoning process and the ground truth by providing the model with prior knowledge, yet it is prone to post-hoc rationalization where the model may merely craft a superficial justification to match the known answer instead of performing genuine logical deduction. In Section~\ref{rq:sft_data_curation}, we will compare and analyze the impact of these two SFT data curation methods on model performance.

\subsubsection{LLM-Judge Filtering Protocol}\label{sec:filter}

When performing rejection sampling on teacher CoTs, LLM-as-a-Judge is employed to serve as the filtering mechanism. Specifically, for each candidate response, the judge LLM determines whether the model analysis has correctly identified the target vulnerability, or correctly confirmed its absence, by judging whether the teacher LLM's reasoning-level root-cause analysis aligns with the provided ground truth vulnerability information (more details can be found in Appendix~\ref{sec:reasoning}). To mitigate the impact of LLM non-determinism and the inherent difficulty of detecting certain vulnerability samples, the teacher LLM is configured to generate eight candidate responses for each VD query. Ultimately, only those queries with at least one correct response are retained to ensure the overall quality of the SFT dataset.

\subsection{Off-policy Preference Optimization}\label{sec:preference_optimization}

\begin{figure}[h]\centering
\includegraphics[width=0.7\columnwidth]{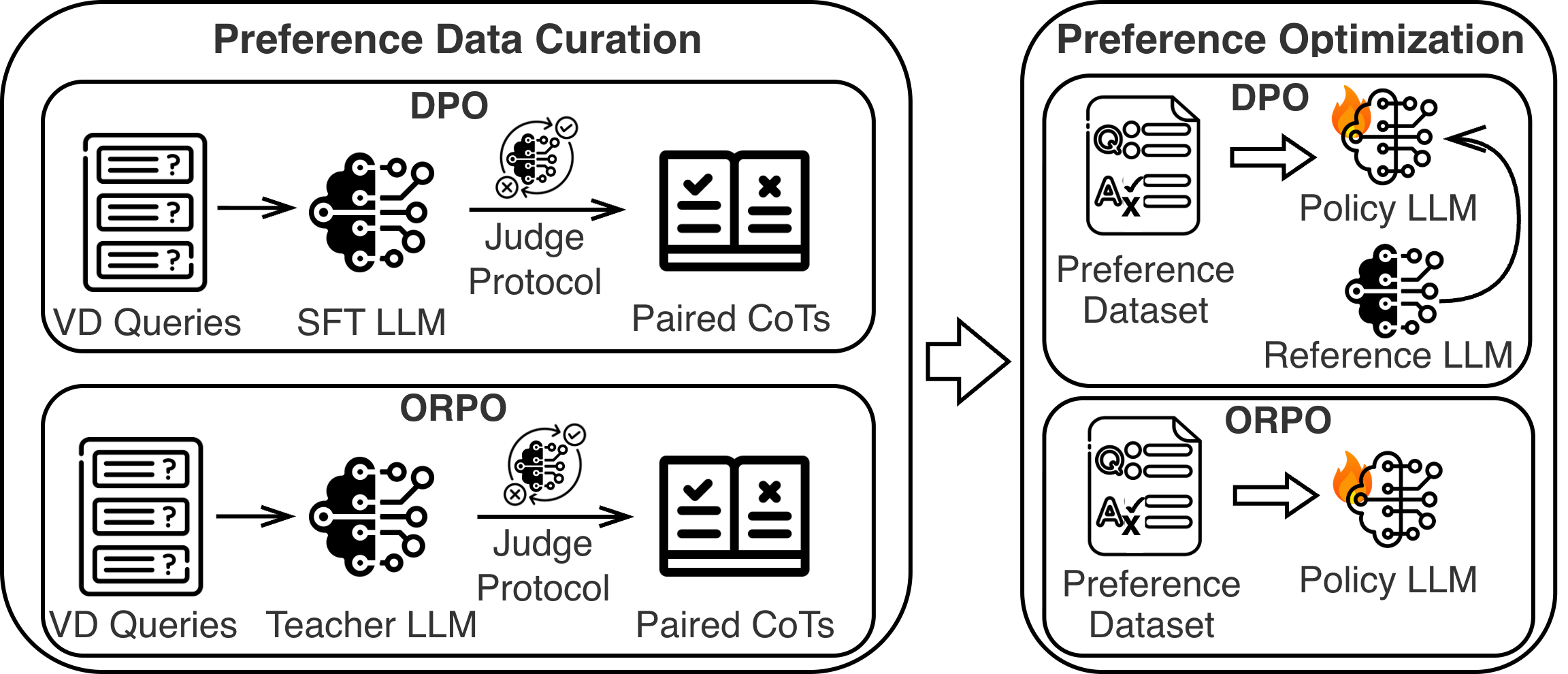}
\caption{Training Pipeline for Preference Optimization} \label{fig:preference}
\end{figure}

Figure~\ref{fig:preference} shows the post-training pipelines for off-policy preference optimization algorithms, DPO and ORPO.

\subsubsection{Preference Data Curation}\label{sec:preference_data_curation}

In DPO, the preference dataset is typically constructed by sampling paired CoTs directly from the SFT LLM to minimize the distribution shift between the SFT LLM and the policy LLM optimized in the subsequent preference optimization process. Specifically, we generate eight candidate responses per query from the SFT LLM. These responses are then processed through the judge protocol (detailed in Section~\ref{sec:filter}) to pair preferred and dispreferred responses from the same query, thereby building the preference dataset. Since ORPO integrates SFT and preference optimization into Equation~\ref{eq:orpo}, its preference dataset is typically generated by a teacher LLM.

\subsubsection{Off-policy Optimization}

DPO typically begins with an SFT LLM as its starting point, which also serves as the reference LLM. It then directly optimizes the policy LLM under a Bradley-Terry-based preference objective, which allows for implicit reward modeling without a separate training stage. ORPO further evolves this process by unifying the SFT and preference optimization stages into a single objective, further eliminating the need for a reference model.

\subsection{On-policy Reinforcement Learning}\label{sec:rl}

\begin{figure}[h]\centering
\includegraphics[width=0.7\columnwidth]{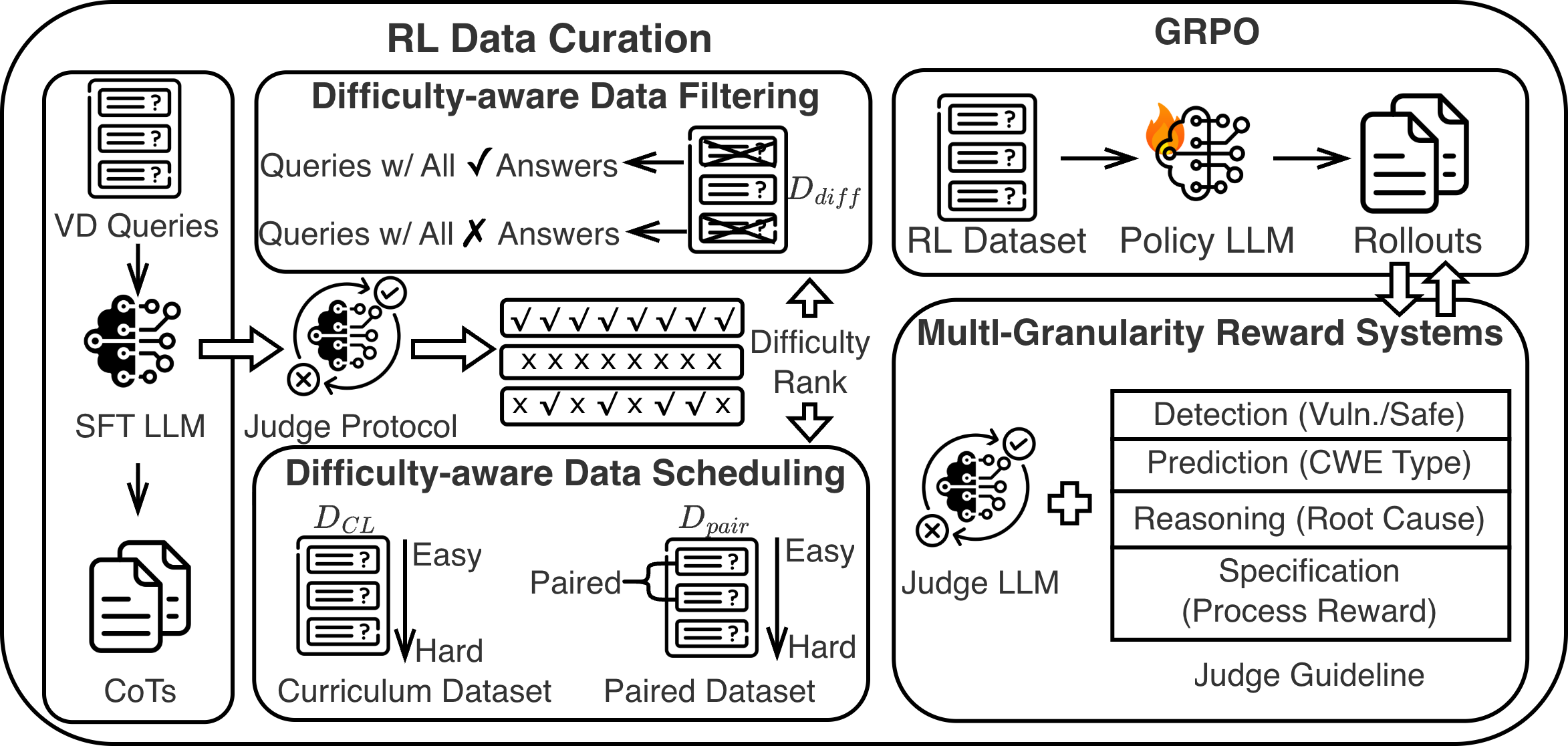}
\caption{Training Pipeline for Reinforcement Learning} \label{fig:rl}
\end{figure}

\subsubsection{RL Data Curation}

In the on-policy RL stage, the difficulty of the vulnerability queries in the training data determines reward sparsity, which in turn dictates the effectiveness and efficiency of RL training. Particularly in VD, the model's performance varies significantly across different vulnerability types (e.g., memory-related vs. complex logic vulnerabilities) and even within the same category when presented in different code structures (e.g., function-level vs. inter-procedural vulnerabilities). Consequently, the curation and organization of VD data are critical factors influencing RL training. Motivated by these insights, we propose the following difficulty-aware data filtering and scheduling schemes, as shown in Figure~\ref{fig:rl}, whose impact on RL training will be analyzed in Section~\ref{rq:data_difficulty}.

\noindent\textbf{Difficulty-aware Data Filtering.} In a full-data setting, when a group of responses for extremely simple or difficult problems yields identical group rewards $\{r_i\}^G_{i=1}$, the group-relative reward normalization in GRPO results in a group advantage of zero for all samples (i.e., $\hat{A}_{i,t} = \frac{r_i - \text{mean}(\{r_i\}^G_{i=1})}{\text{std}(\{r_i\}^G_{i=1})}=0$). This inhibits effective parameter updates, as gradient signal effectively vanishes for these samples. 

\begin{wrapfigure}{r}{0.5\columnwidth}
\centering
\includegraphics[width=0.48\columnwidth]{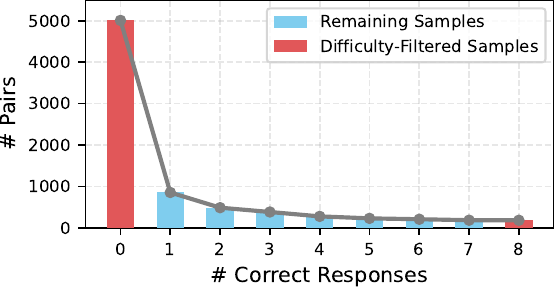}
\caption{Difficulty Distribution of $\mathcal{D}_{diff}$ on SFT LLM} \label{fig:difficulty_distribution}
\end{wrapfigure}

To address this, we propose a difficulty-aware data filtering strategy. We quantify the difficulty of vulnerability queries using the SFT LLM's Pairwise Pass@1. Specifically, for each pair, comprising a vulnerable code snippet and its corresponding patched code, the SFT LLM generates eight response pairs. A response pair is deemed correct only if the model correctly identifies the target vulnerability in the vulnerable code and recognizes its absence in the patched version (i.e., achieving both a true positive and a true negative). This pairwise constraint prevents the filtered dataset from being dominated by a single class due to inherent model prediction bias (e.g., a tendency toward false positives/negatives). We then construct $\mathcal{D}_{\text{diff}}$ by excluding pairs with Pairwise Pass@1 of 0 or 1 (all eight response pairs are incorrect or correct). Figure~\ref{fig:difficulty_distribution} displays the difficulty distribution of $\mathcal{D}_{\text{diff}}$ on the SFT LLM. 

\noindent\textbf{Difficulty-aware Data Scheduling.} Traditional random sampling often introduces high variance in batch difficulty and creates imbalances between vulnerable and patched samples, potentially destabilizing RL training. To address this, we propose two difficulty-aware data scheduling strategies:
\begin{itemize}[leftmargin=10pt]
\item \textit{Curriculum Learning}: We sort the VD queries by Pairwise Pass@1 in descending order to construct $\mathcal{D}_{\text{CL}}$. This ensures a gradual increase in difficulty across batches, allowing the model to learn simple vulnerability patterns before tackling complex samples.
\item \textit{Pairwise Scheduling}: Building on $\mathcal{D}_{\text{CL}}$, we further construct $\mathcal{D}_{\text{pair}}$, ensuring that a vulnerable sample and its patched version are always included in the same training batch while maintaining the gradual progression of difficulty across batches.
\end{itemize}

\subsubsection{On-policy Optimization}

As illustrated in Figure~\ref{fig:rl}, GRPO typically begins with a cold-start stage to obtain an SFT LLM as the initial policy LLM. Subsequently, in each GRPO iteration, RL data processed through difficulty-aware data curation is fed into the policy LLM to generate online rollouts. These group rollouts are then scored by a reward system involving a judge LLM, as introduced in Section~\ref{sec:mas}, providing reward feedback to guide on-policy LLM optimization.

\subsubsection{Multi-granularity Reward Systems}\label{sec:mas}

In RL, the reward signal guides model training, making reward design critical to overall effectiveness. While rule-based rewards, such as unit-test verifiers for resolving GitHub issues, are effective for general coding tasks~\citep{67,68,69,79}, software security presents unique challenges. The difficulty of reproducing vulnerabilities, coupled with their strict dependency on specific software versions, complicates the large-scale acquisition of verifiable VD samples and the provisioning of their necessary execution environments.

Early VD efforts rely on binary comparisons for evaluation~\citep{60}. This coarse-grained approach often results in unreliable credit assignment, for instance, rewarding a model that guesses correctly despite flawed reasoning, which misleads the RL training process. Since most VD datasets are sourced from public databases like the National Vulnerability Database (NVD)~\citep{84}, we can leverage this metadata as a proxy for reward assessment. Consequently, we propose multi-granularity reward systems spanning detection, prediction, reasoning, and specification levels to systematically explore their impact on RL as follows:

\noindent\textbf{Detection-based Reward.} This baseline reward measures the accuracy of binary detection. A model receives a reward of $+1.0$ for a correct binary label and a penalty of $-1.0$ otherwise.

\noindent\textbf{Prediction-based Reward.} This level evaluates the model's ability to predict the specific CWE type. To account for the hierarchical nature of the CWE, we avoid a naive exact-match approach, which often leads to biased credit assignment. Following the mapping methodology in~\citep{61,62}, a prediction is deemed correct, yielding a $+1.0$ reward, if the predicted CWE ID and the ground truth share a direct parent-child relationship in the CWE taxonomy; otherwise, it receives a penalty of $-1.0$. More details about CWE matching can be found in Appendix~\ref{sec:cwe_match}.

\noindent\textbf{Reasoning-based Reward.} To move beyond superficial lexical overlap, we employ an LLM-as-a-Judge to assess the semantic integrity of the model's vulnerability analysis. The judge LLM evaluates whether the model's root cause analysis aligns with the ground truth (i.e., CVE description, commit message, and patch diff) based on the judge guideline detailed in Appendix~\ref{sec:reasoning}. The model receives a $+1.0$ reward if so, and a $-1.0$ penalty otherwise. We argue that reasoning-based rewards are more robust than prediction-based ones, as they mitigate reward hacking where models might correctly guess a common vulnerability type (e.g., buffer overflow) through flawed logic.

\noindent\textbf{Specification-based Reward.} While aforementioned systems focus on outcome rewards, they typically fail to (1) distinguish between reasoning paths of varying procedural quality that reach the same conclusion, and (2) provide sample-specific granularity beyond general judge guidelines. To solve these, we propose a specification-based process reward system. For each sample, an LLM pre-generates a tailored specification derived from ground truth. This specification defines three dimensions for assessing vulnerability analysis:
\begin{itemize}[leftmargin=10pt,itemsep=2pt, parsep=2pt]
\item \textit{Prediction Accuracy}: whether the analysis correctly predicts the vulnerability type (e.g., a buffer overflow)
\item \textit{Localization Precision}: whether the analysis correctly localizes the exact vulnerability-related code snippet (or a semantically equivalent variation)
\item \textit{Reasoning Quality}: whether the analysis clearly explains both the root cause (e.g., a missing check) and the resulting consequence (e.g., an overflow)
\end{itemize}

\begin{wrapfigure}{r}{0.5\columnwidth}
\centering
\includegraphics[width=0.48\columnwidth]
{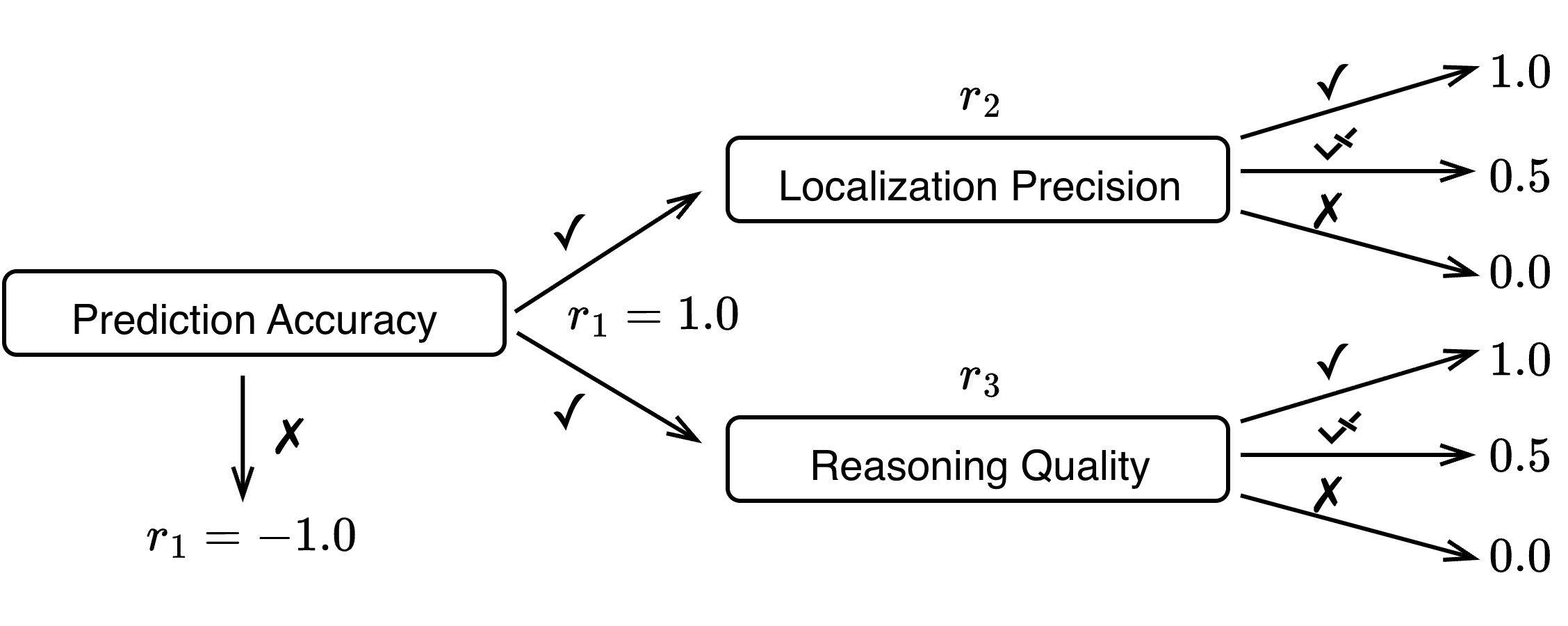}
\caption{Rubric for Specification-based Reward Assignment} \label{fig:reward_map}
\end{wrapfigure}

Figure~\ref{fig:reward_map} shows the detailed rubric for specification-based reward assignment. Prediction accuracy serves as a correctness gate: if the model correctly predicts the vulnerability type, it receives a reward of $r_1=1.0$; otherwise, it is assigned a penalty of $r_1=-1.0$. The other two dimensions, localization precision and reasoning quality, act as incremental rewards, each taking values of correct ($1.0$), partially correct ($0.5$), or incorrect ($0.0$), to differentiate quality of reasoning chains that correctly predict the vulnerability type.

To normalize the total reward $R$, we assign weights of $0.4$, $0.2$, and $0.4$ (i.e., 2:1:2) to prediction accuracy, localization precision, and reasoning quality, respectively, $R$ can be computed as a weighted average of these three dimensions:
\begin{equation}
R = 0.4 \cdot r_1 + 0.2 \cdot r_2 + 0.4 \cdot r_3.
\end{equation}

This weighting is motivated by our empirical observation that many model outputs correctly predict the vulnerability type but fail to localize the relevant code snippet, and even when localization is correct, the root-cause analysis may still be inconsistent with the ground-truth vulnerability. Overall, this specification-based process reward design enables the judge LLM to distinguish between correct and incorrect vulnerability reasoning, discern quality differences among correct reasoning chains, and reduce the stochasticity inherent in general guideline following. Detailed prompts for sample-level specification generation and judgment are provided in Appendix~\ref{sec:generation}-\ref{sec:specification}. Section~\ref{rq:reward} conducts experimental comparisons across four reward granularities.

\subsection{Evaluation Protocol}\label{sec:protocol}
The accuracy of evaluation determines the subsequent optimization direction of the policy model and influences the estimation of a VD LLM's practical capability in real-world scenarios. Therefore, the evaluation protocol is a critical component of the post-training pipeline for LLM-based VD. We introduce the following three evaluation methods, ranging from coarse-grained to fine-grained, and will discuss their impact on evaluation results in Section~\ref{rq:sankey}:
\begin{itemize}[leftmargin=10pt,,itemsep=2pt, parsep=2pt]
\item \textit{Binary-label Matching:} A detection is considered a True Positive (TP) or True Negative (TN) when the model's predicted binary label matches the ground truth; otherwise, it is classified as a False Positive (FP) or False Negative (FN).
\item \textit{CWE Matching:} For vulnerable samples, a prediction is a TP if the model's predicted CWE has a direct parent-child relationship with the ground truth; otherwise, it is an FN. For patched versions where vulnerabilities have been fixed, the prediction is considered a TN if the model identifies the sample as benign or it identifies unknown CWE types; otherwise, it is an FP.
\item \textit{Root-cause Matching (default evaluation method):} For vulnerable samples, a vulnerability reasoning is considered a TP only when the model's root cause analysis aligns with the ground truth vulnerability information; otherwise, it is an FN. For patched versions, the vulnerability reasoning is considered a TN if the model identifies the sample as safe or it identifies unknown vulnerabilities; otherwise, it is an FP.
\end{itemize}

It should be noted that existing VD datasets are typically composed of vulnerable code and its patched version extracted from commits addressing specific target CVEs. For each VD sample, we only know whether it contains the specific target vulnerability, while the presence of other vulnerabilities remains unknown. Consequently, the evaluation method adopted in this work treats cases where the model identifies other unknown vulnerabilities in samples where the target CVE has been patched as TN rather than FP. This conservative approach reports the upper bound of the model's precision. Addressing the inherent challenges of VD dataset quality and more precise evaluation remains a subject for future research.

Finally, as the reliability of root-cause matching depends on the LLM-as-a-Judge, Section~\ref{rq:human} will compare different judge LLMs with human-verified results to measure potential bias.
\section{Experimental Setup}
\label{sec:data_split}

\begin{table}[h]
\caption{Dataset Statistics}
\label{tab:statistics}
\centering
\resizebox{0.7\columnwidth}{!}{
\renewcommand{\arraystretch}{1.0}
\begin{tabular}{ccc|cccc}
\hline
\textbf{\#Samples} & \textbf{\#Projects} & \textbf{\#CWEs} & \textbf{\#Tokens} & \textbf{Function} & \textbf{Context} & \textbf{Total Input} \\ \hline
\multirow{3}{*}{\begin{tabular}[c]{@{}c@{}}Train: 15594\\ Val.: 1968\\ Test: 1970\end{tabular}} & \multirow{3}{*}{775} & \multirow{3}{*}{179} & Min & 27 & 0 & 296 \\
 &  &  & Mean & 729 & 1270 & 2242 \\
 &  &  & Max & 14888 & 14311 & 16292 \\ \hline
\end{tabular}
}
\end{table}

\noindent\textbf{Dataset Partitioning.} To avoid data leakage, we partition the dataset introduced in Section~\ref{sec:pre} into training, validation, and test sets according to an 8:1:1 ratio based on commit dates. Given the pair-based nature of our dataset, a balanced distribution is critical to mitigate bias and ensure the effectiveness and fairness of the trained model. Table~\ref{tab:statistics} presents the statistics of the dataset.

\vspace{0.3em}
\noindent\textbf{Models.} We adopt Qwen3-4B as our backbone LLM for its optimal balance between parameter efficiency and advanced reasoning capabilities in coding tasks. To construct the reasoning datasets detailed in Sections~\ref{sec:sft_data_curation} and~\ref{sec:preference_data_curation}, we employ DeepSeek-R1-0528 as the teacher LLM. For reward assessment during RL training, as well as in our LLM-judge-based filtering and evaluation protocols, GPT-OSS-120B, configured with high reasoning effort, is selected as the default judge LLM and sample specification generator. Furthermore, we present a comparative analysis of several general-purpose LLMs and specialized VD LLMs under zero-shot settings, along with a comprehensive evaluation of various post-training methods built upon Qwen3-4B, as detailed in Section~\ref{rq:0shot}.

\vspace{0.3em}
\noindent\textbf{Metrics.}
As described in Section~\ref{sec:protocol}, our VD dataset is constructed from CVE patch commits, where each vulnerable code snippet has a corresponding patched version. An effective VD model should not only correctly identify the target vulnerability in the vulnerable code, but also recognize its absence in the patched version.
Accordingly, we propose Pairwise Pass@k (P-Pass@k) as the primary evaluation metric. For each vulnerability-patch query pair, we sample eight response pairs, where each pair consists of one response for the vulnerable code and one response for its patched counterpart. Let $p_{i,j} \in \{0,1\}$ denote whether the $j$-th response pair for the query pair $i$ is jointly correct, i.e., the model correctly identifies the target vulnerability in the vulnerable version and recognizes its absence in the patched version. Then, the P-Pass@k metrics are defined as: $\text{P-Pass@1} = \frac{1}{Nk}\sum\limits_{i=1}^{N}\sum\limits_{j=1}^{k} p_{i,j}$, $\text{P-Pass@8} = \frac{1}{N}\sum\limits_{i=1}^{N} \left[ 1 - \prod\limits_{j=1}^{k} (1 - p_{i,j}) \right]$.
We further introduce three pairwise consistency metrics, P-V, P-B, and P-R~\citep{17}, 
to characterize model behavior on vulnerability-patch pairs. Note that we omit P-C since it is equivalent to P-Pass@1. Specifically, P-V denotes cases where the model identifies the target vulnerability in both versions; P-B denotes cases where the model predicts that the target vulnerability is absent in both versions; and P-R denotes cases where the predictions for both versions contradict the ground truth. In addition, we report the F1 score based on standard sample-level predictions: $\text{F1} = \frac{2TP}{2TP + FP + FN}$.

\vspace{0.3em}
\noindent\textbf{Implementation.}
Our post-training framework for LLM-based VD is built upon the Hugging Face TRL library, integrating optimization techniques such as DeepSpeed and vLLM to enhance training and inference efficiency. All SFT and preference optimization models are trained on four NVIDIA A100-SXM4-80GB GPUs (\textasciitilde16 GPU-hours in total). For GRPO, we employ a distributed RL setup across two nodes totaling eight GPUs (\textasciitilde384 GPU-hours in total), with two additional GPUs hosting a vLLM server to facilitate high-throughput online rollout generation and offline inference (\textasciitilde8 GPU-hours in total). Regarding hyperparameters, for cold-start SFT, we use a learning rate of $1e-5$ with a linear schedule, a warmup ratio of $0.1$, a total batch size of $32$, and $8$ gradient accumulation steps over $5$ epochs. For DPO, ORPO, and GRPO, the learning rate is adjusted to $1e-6$. Regarding the $\beta$ hyperparameter in Table~\ref{tab:obj}, we perform a grid search across $\{0.01, 0.1, 1\}$ for DPO and ORPO, reporting the optimal performance at $\beta=0.1$. For GRPO, following the common practice in~\citep{85,86}, we set $\beta=0$ to maximize training efficiency. All other hyperparameters remain consistent with the SFT stage. For zero-shot prompting, we utilize default model configurations. To ensure comprehensive contextual reasoning, we set the maximum sequence length to 32,768 tokens, providing a strategic balance between accommodating long code contexts and reserving sufficient space for reasoning steps.
\section{Experimental Evaluation and Results}

We present our evaluation results and provide practical insights structured around the following research questions (RQs). Unless otherwise specified, each section's experiments build upon the optimal configurations identified in earlier sections. 

\subsection{RQ1: How do different SFT data curation methods impact VD LLM performance?}~\label{rq:sft_data_curation}

\begin{wrapfigure}{r}{0.45\columnwidth}
\centering
\includegraphics[width=0.4\columnwidth]{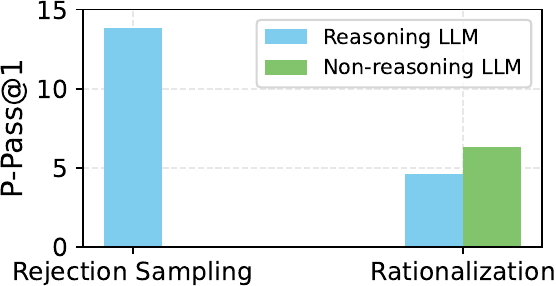}
\caption{Comparison of SFT Data Curation Methods} \label{fig:pov_sft_data_curation}
\end{wrapfigure}

Rejection sampling has become a common practice for constructing LLM post-training datasets~\citep{77,83,85,86}. However, the difficulty of VD makes it challenging to sample a sufficient number of correct vulnerability reasoning traces from LLMs. As a result, existing VD work~\citep{57, 58} adopts rationalization, which injects vulnerability-related ground-truth information into the model input to guide it toward producing correct vulnerability analysis. Therefore, in this section, we compare and analyze the impact of the data curation methods (i.e., rejection sampling vs. rationalization) introduced in Section~\ref{sec:sft_data_curation} on the performance of SFT LLMs.

As shown in Figure~\ref{fig:pov_sft_data_curation}, the SFT LLM trained on the dataset curated via rejection sampling significantly outperforms the one trained via rationalization, achieving a relative P-Pass@1 improvement of 201.7\%. We observe that despite the test data consisting solely of raw code, the rationalization-based SFT LLM suffers from hallucinations. Specifically, it memorizes vulnerability knowledge learned during training, such as hallucinating CVE IDs or vulnerability types that are not actually present in the test input, rather than learning to derive the vulnerability from the code itself. This distribution shift between the teacher's privileged context and the student's restricted context leads to a failure in learning the underlying reasoning logic.

To mitigate this issue, we exclude the reasoning content within the \texttt{<think>} tags generated by the teacher model while retaining the answers that do not contain ground-truth information. We then train an additional non-reasoning LLM on this reprocessed SFT dataset. Figure~\ref{fig:pov_sft_data_curation} shows that while the non-reasoning LLM achieves a relative P-Pass@1 increase of 37.0\% compared to the reasoning one, its performance remains substantially lower than that of the SFT LLM trained on the rejection sampling dataset. These results indicate that the rationalization-based data curation method severely impairs the effective training of models during the SFT stage. A potential solution involves filtering information from rationalization-generated datasets (e.g., using an auxiliary LLM).

\begin{tcolorbox}[colback=orange!10, colframe=orange!100, boxrule=1pt, left=1mm, right=1mm, top=1mm, bottom=1mm]
\textbf{Answer-1:} Rejection sampling is significantly more effective than rationalization for SFT data curation. While rationalization causes models to hallucinate CVE IDs or vulnerability types that are not actually present in the test input, rather than learning from the code, rejection sampling ensures higher data quality. Even after removing reasoning parts to mitigate hallucinations, rationalization-based models still underperform.
\end{tcolorbox}

\subsection{RQ2: How does the extent of SFT influence the performance of VD LLMs subsequently trained through preference optimization?}\label{rq:sft2dpo}

\begin{figure}[h]
\centering
\includegraphics[width=0.8\columnwidth]{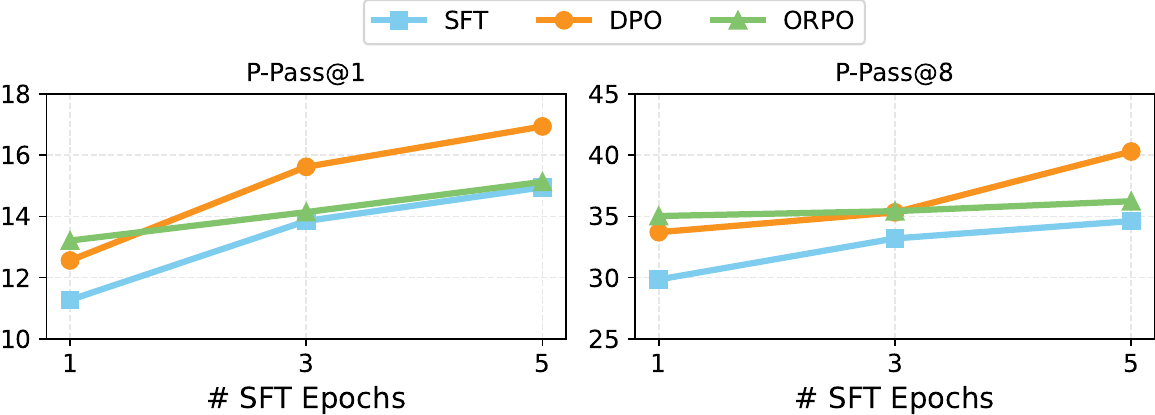}
\caption{Impact of SFT on Preference Optimization} \label{fig:pov_sft2dpo}
\end{figure}

Post-training work in the general domain~\citep{80,81,88} typically uses lightly trained SFT models (e.g., 1-2 epochs) as the initialization for the subsequent preference optimization stage. This minimizes distribution shift and provides a baseline for distinguishing between preferred and dispreferred responses.  However, it remains unclear how the degree of SFT training affects off-policy preference optimization, especially in VD tasks, which motivates us to investigate the performance of SFT checkpoints across different epochs and their impact on DPO and ORPO. Although ORPO unifies SFT and preference optimization, we use the SFT model as the starting point for both DPO and ORPO to ensure a fair comparison, following~\citep{88}.

The results in Figure~\ref{fig:pov_sft2dpo} show that as SFT epochs increase, the performance of the policy LLM improves, leading to a consistent performance gain in both DPO and ORPO. Specifically, the P-Pass@1 score of the DPO model initialized from a 5-epoch SFT checkpoint relatively improves by 34.9\% compared to a 1-epoch initialization, whereas the ORPO achieves a relative improvement of 14.5\%. Notably, while ORPO outperforms DPO when initialized with a 1-epoch SFT checkpoint, DPO's P-Pass@1 and P-Pass@8 scales more significantly with more SFT training and eventually surpasses ORPO. This suggests that DPO is more dependent on a high-quality SFT foundation to effectively differentiate vulnerability-patch pairs. In contrast, ORPO demonstrates greater robustness and less dependency on the initial SFT stage, highlighting its effectiveness in unifying SFT and preference optimization.

\begin{tcolorbox}[colback=orange!10, colframe=orange!100, boxrule=1pt, left=1mm, right=1mm, top=1mm, bottom=1mm]
\textbf{Answer-2:} Extensive SFT significantly boosts the performance of models subsequently trained through preference optimization, particularly DPO. While ORPO is more robust and less dependent on the initialization stage, DPO scales more effectively with a higher-score SFT foundation, eventually surpassing ORPO as SFT epochs increase.
\end{tcolorbox}

\subsection{RQ3: How does the extent of SFT influence the performance of VD LLMs subsequently trained through on-policy RL?}

\begin{figure}[h]\centering
\includegraphics[width=0.8\columnwidth]{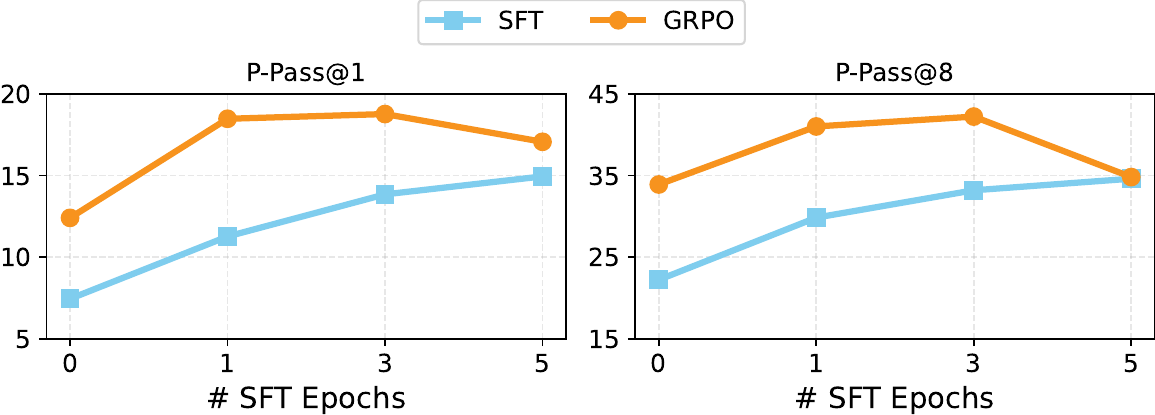}
\caption{Impact of SFT on On-Policy RL} \label{fig:pov_sft2grpo}
\end{figure}

Cold-start SFT is essential for LLM post-training, as seen in models like DeepSeek-R1~\citep{77}. It builds the necessary foundations for downstream tasks, increasing the probability of the model generating correct rollouts during the subsequent RL stage and mitigating the impact of sparse rewards on training. However, there are conflicting views between SFT and RL. Common practice typically selects models with better post-SFT performance as the starting point for RL training~\citep{91,92}, assuming this will consistently yield superior final outcomes after RL. In contrast, recent studies~\citep{93,94} suggest that SFT models with high scores are not necessarily optimal initializations for RL. This motivates us to investigate how to balance SFT and RL optimization by examining the impact of SFT intensity on the on-policy RL algorithm GRPO in VD tasks. 

Figure~\ref{fig:pov_sft2grpo} shows model performance across different SFT checkpoints and the relative gains achieved after GRPO. The results confirm that the GRPO model without cold-start SFT performs significantly worse than others, confirming that SFT remains a critical initialization stage to RL training for VD.
However, in contrast to the continuous performance improvements seen in preference optimization with stronger SFT initialization, the relative performance gains of GRPO models over their SFT baselines decrease significantly as the number of SFT training epochs increases, eventually leading to a negative impact. Specifically, the P-Pass@1 score of the GRPO model trained on the epoch=5 SFT checkpoint decreases by 9.1\% relative to that at epoch=3. Furthermore, the sharp decline in P-Pass@8 indicates that excessive SFT restricts the model exploration during RL training. This prevents the model from sampling diverse responses, ultimately leading to lower accuracy.

\begin{figure}[h]
\centering
\includegraphics[width=\columnwidth]{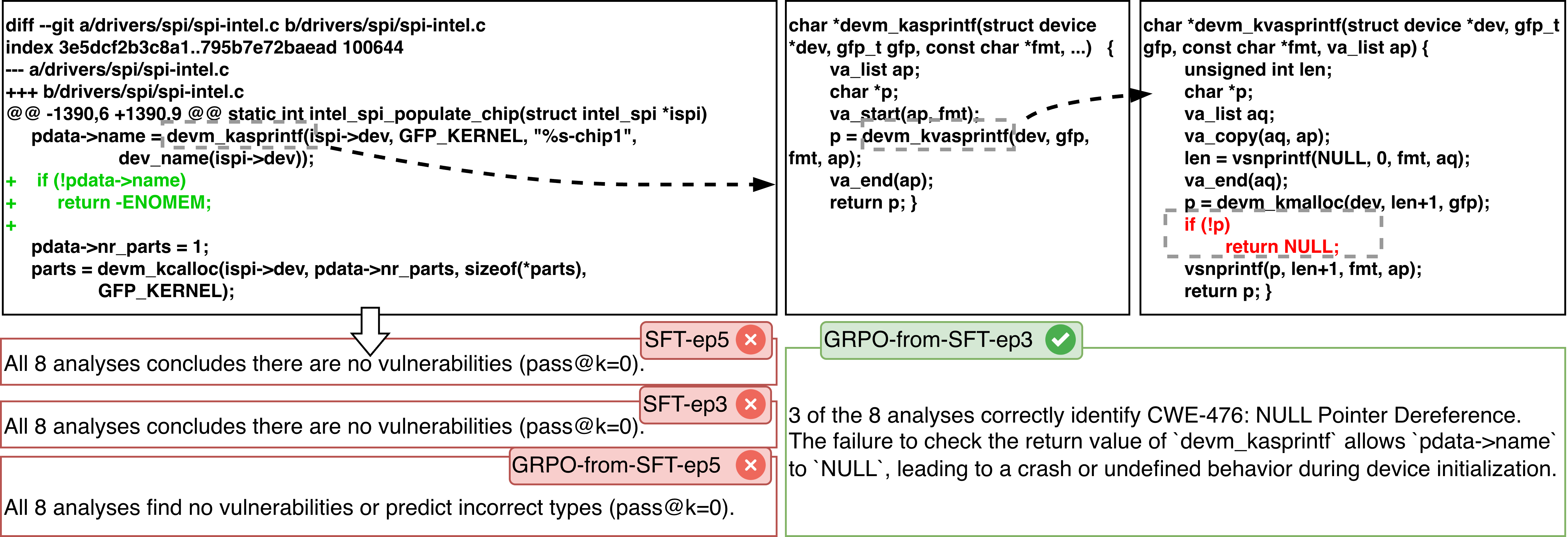}
\caption{CVE-2024-46769 - NULL Pointer Dereference} \label{fig:null}
\end{figure}

\noindent\textbf{Case Study.} 
Figure~\ref{fig:null} illustrates a concrete NULL pointer dereference vulnerability, CVE-2024-46769. The vulnerability is caused by a missing NULL check after calling \texttt{devm\_kasprintf}. This function internally relies on \texttt{devm\_kmalloc} for memory allocation, which may return NULL on failure. In the vulnerable code, the returned pointer is directly assigned to \texttt{pdata->name} without validation. If allocation fails, \texttt{pdata->name} becomes NULL, and subsequent usage leads to a NULL pointer dereference (CWE-476), potentially causing a crash or undefined behavior during device initialization. 

The detection results of different models on this vulnerable function further validate that:
(1) excessive SFT (epoch=5) suppresses exploration in the RL stage, preventing the GRPO model from discovering the vulnerability;
(2) with moderate SFT (epoch=3), RL can effectively refine reasoning trajectories and explore a broader space, enabling the GRPO model to consider edge cases such as memory allocation failure and ultimately sample correct rollouts.

\begin{tcolorbox}[colback=orange!10, colframe=orange!100, boxrule=1pt, left=1mm, right=1mm, top=1mm, bottom=1mm]
\textbf{Answer-3:} Cold-start SFT is essential for GRPO to succeed in VD, but its intensity must be carefully balanced. While moderate SFT provides a necessary foundation, excessive SFT prevents the model from exploring diverse reasoning paths, thereby decreasing P-Pass@1 and P-Pass@8.
\end{tcolorbox}

\subsection{RQ4: How do data filtering and scheduling influence the effectiveness of RL training?}\label{rq:data_difficulty}

\begin{wraptable}{r}{0.6\columnwidth}
\caption{Impact of RL Data Curation on Model Performance}
\label{tab:rl_data_curation}
\centering
\resizebox{0.58\columnwidth}{!}{
\renewcommand{\arraystretch}{1.0}
\begin{tabular}{ccccc}
\toprule
\textbf{Data Filtering} & \textbf{Data Scheduling} & \textbf{P-Pass@1} & \textbf{P-Pass@8} & \textbf{F1} \\ \midrule
Full Participation & \begin{tabular}[c]{@{}c@{}}Random\\ Curriculum~$\mathcal{D}_{CL}$\\ Paired~$\mathcal{D}_{pair}$\end{tabular} & \begin{tabular}[c]{@{}c@{}}18.77\\ 18.78\\ 18.90\end{tabular} & \begin{tabular}[c]{@{}c@{}}42.43\\ 41.42\\ 41.32\end{tabular} & \begin{tabular}[c]{@{}c@{}}34.36\\ 34.95\\ 34.50\end{tabular} \\ \midrule
Difficulty Filtering & \begin{tabular}[c]{@{}c@{}}Random\\ Curriculum~$\mathcal{D}_{CL}$\\ Paired~$\mathcal{D}_{pair}$\end{tabular} & \begin{tabular}[c]{@{}c@{}}16.02\\ 15.71\\ 17.07\end{tabular} & \begin{tabular}[c]{@{}c@{}}36.35\\ 35.23\\ 38.88\end{tabular} & \begin{tabular}[c]{@{}c@{}}29.83\\ 29.78\\ 31.48\end{tabular} \\ \bottomrule
\end{tabular}
}
\end{wraptable}

Numerous studies have demonstrated the benefits of curriculum learning for LLM-based RL in general code tasks~\citep{89,90}. Given the complexity and diversity of vulnerability patterns, LLM performance varies significantly across different samples. Motivated by this, this section investigates the impact of various RL data curation strategies on RL training. Table~\ref{tab:rl_data_curation} reports the performance of GRPO models trained under two data filtering settings and three scheduling methods.

\noindent\textbf{Impact of Data Filtering.} Our results indicate that filtering out extremely simple and difficult training data results in a performance drop compared to the full dataset; under random scheduling, we observe relative decreases of 14.65\% in P-Pass@1, 14.33\% in P-Pass@8, and 13.18\% in F1. This reveals that difficulty-aware data filtering in VD risks discarding a large amount of potentially valuable training data that contributes to model performance. However, even the GRPO model trained on the filtered dataset achieves a relative P-Pass@1 improvement of 15.7\% over the initial checkpoint, indicating that the choice of filtering should be a strategic trade-off between computational efficiency and peak performance.

\noindent\textbf{Impact of Data Scheduling.} Under both filtering settings, curriculum learning fails to bring performance gains over the baseline. This is likely due to the inherent difficulty of VD compared to general coding tasks. As shown in Figure~\ref{fig:difficulty_distribution}, the difficulty distribution of training data is heavily skewed toward the extreme difficulty. Consequently, the model struggles to sample correct rollouts for a large proportion of hard problems in later training stages, even after mastering simpler ones, thereby depriving it of the positive feedback necessary for effective RL. In contrast, GRPO models trained on paired datasets consistently outperform those on original datasets. This effect may arise because paired data scheduling allows the model to learn the subtle differences between vulnerabilities and their corresponding patches within the same batch. This strengthens the model's understanding of subtle vulnerability patterns, thereby enhancing overall performance.

\begin{tcolorbox}[colback=orange!10, colframe=orange!100, boxrule=1pt, left=1mm, right=1mm, top=1mm, bottom=1mm]
\textbf{Answer-4:} Filtering extreme cases in VD risks a performance drop by discarding a large amount of potentially valuable training signals. Traditional curriculum learning is ineffective in VD due to a difficulty distribution skewed toward hard questions, which starves the model of positive feedback. However, paired data scheduling helps the model learn subtle nuances between vulnerabilities and patches, thereby improving performance.
\end{tcolorbox}

\subsection{RQ5: How does the granularity of reward systems impact the efficiency of RL?}\label{rq:reward}

\begin{wrapfigure}{r}{0.55\columnwidth}
\centering
\includegraphics[width=0.52\columnwidth]{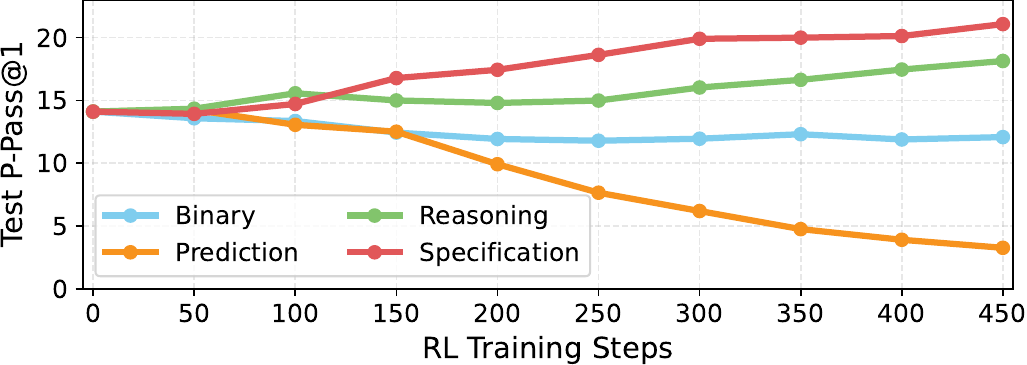}
\caption{Impact of Reward Granularities on RL Training} \label{fig:granularity}
\end{wrapfigure}

Prior VD work typically evaluates model performance based on the accuracy of binary detection~\citep{58,60} or CWE prediction~\citep{55, 61}. This has further inspired subsequent RL-for-VD approaches~\citep{60} to use the correctness of binary detection as the reward signal for guiding RL training. However, correctly identifying a vulnerability requires not only predicting the correct binary label, but also accurately determining the vulnerability type and its root cause. Coarse-grained reward design can therefore negatively affect model updates during the RL stage. This motivates our investigation in this section into how different levels of reward granularity affect GRPO training in the RL stage.

As illustrated in Figure~\ref{fig:granularity}, as RL training progresses, the GRPO model trained under the prediction-based reward system experiences a sharp performance drop. This degradation occurs because correctly predicting the vulnerability type is much harder than predicting that the target code is safe. As a result, the model learns to take a shortcut during RL training by predicting nearly all samples as safe, leading to reward hacking. Similarly, the GRPO model trained under the detection-based reward system even performs worse than the initial model obtained from the cold-start SFT stage. This is because rewards based on binary-label correctness often coincidentally reinforce outputs where the model guesses the correct label through flawed reasoning, causing model updates to deviate from the intended direction.

In contrast, the fine-grained reasoning-based and specification-based reward systems improve the stability and efficiency of RL training. The effectiveness of the reasoning-based reward stems from the fact that accurate root cause analysis requires the model to precisely locate suspicious vulnerability sources and sinks, conduct comprehensive reasoning, and accurately identify the vulnerability. Guided by this fine-grained reward, the judge LLM can effectively identify model outputs that have correct labels but incorrect root-cause analysis. Notably, although the specification-based reward system further improves RL training efficiency by refining reward assessment to the sample level, practical applications must account for the additional overhead introduced by designing and generating sample specifications, as well as the sensitive analysis on RL training, which we leave to future work. Therefore, to ensure the generalizability of this work, we select the reasoning-based reward system as our default setup.

\begin{tcolorbox}[colback=orange!10, colframe=orange!100, boxrule=1pt, left=1mm, right=1mm, top=1mm, bottom=1mm]
\textbf{Answer-5:} Traditional detection or prediction-based rewards in RL for VD often fail because they encourage reward hacking, where models guess labels without correct reasoning. In contrast, reasoning-based and specification-based rewards improve training stability and efficiency by validating the root cause analysis, with the reasoning-based reward striking the balance between model performance and implementation overhead.
\end{tcolorbox}

\subsection{RQ6: How do zero-shot LLMs and post-training methods perform in VD?}\label{rq:0shot}

\begin{table}[h]
\caption{Performance Comparison between Prompt-based and Post-training Methods in VD}
\label{tab:performance_comparison}
\centering
\resizebox{0.8\columnwidth}{!}{
\renewcommand{\arraystretch}{1}
\begin{tabular}{ccccccc}
\toprule
                                 & \textbf{P-Pass@1} & \textbf{P-Pass@8} & \textbf{F1} & \textbf{P-B$\downarrow$} & \textbf{P-V$\downarrow$} & \textbf{P-R$\downarrow$} \\ \hline
\textbf{Zero-Shot Prompting}     &                 &                 &             &                          &                          &                          \\
Qwen3-4B                         & 7.45           & 22.23           & 15.20       & 87.61                    & 1.18                     & 3.76                     \\
R2VUL-7B~\citep{58}             & 3.78           & 16.24           & 14.68       & 81.33                    & 5.32                     & 9.57                     \\
VulnLLM-R-7B~\citep{55}         & 5.08           & 21.73           & 12.18       & 89.61                    & 1.75                     & 3.57                     \\
DeepSeek-R1-0528-Qwen3-8B       & 11.78          & 32.59           & 22.48       & 82.82                    & 1.57                     & 3.83                     \\
gpt-oss-20b                     & 12.98          & 26.09           & 26.18       & 81.62                    & 2.89                     & 2.50                     \\
o4-mini                         & 15.98          & 31.78           & 29.94       & 80.74                    & 2.21                     & 1.08                     \\
Qwen3-235B-A22B                 & 16.26          & 35.43           & 29.86       & 78.55                    & 2.21                     & 2.98                     \\
DeepSeek-R1-0528                & \underline{21.55}          & \underline{42.03}           & \underline{37.32}       & 69.37                    & 3.48                     & 5.61                     \\
DeepSeek-V3.1                   & \textbf{22.50}          & \underline{42.13}           & \textbf{39.64}       & 71.95                    & 3.59                     & 1.95                     \\
gemini-2.5-flash                & \underline{19.66}          & 39.29           & 35.85       & 75.18                    & 3.31                     & 1.85                     \\ \midrule
\textbf{Off-Policy Optimization} &                 &                 &             &                          &                          &                          \\
Qwen3-4B-SFT                    & 14.95          & 34.62           & 28.26       & 79.70                    & 2.39                     & 2.97                     \\
Qwen3-4B-R2VUL~\citep{58}      & 5.36           & 25.99           & 12.26       & 84.87                    & 1.81                     & 7.96                     \\
Qwen3-4B-ReVD~\citep{57}       & 4.63           & 23.25           & 11.28       & 78.19                    & 2.37                     & 14.81                    \\
Qwen3-4B-DPO                    & 16.94          & 40.30           & 31.28       & 76.05                    & 2.89                     & 4.11                     \\
Qwen3-4B-ORPO                   & 15.13          & 36.24           & 28.79       & 77.42                    & 2.94                     & 4.51                     \\ \midrule
\textbf{On-Policy RL}           &                 &                 &             &                          &                          &                          \\
Qwen3-4B-MARCO~\citep{60}      & 11.52          & 27.51           & 22.24       & 85.51                    & 1.36                     & 1.61                     \\
Qwen3-4B-OpenVul               & \underline{21.08}          & \textbf{47.01}           & \underline{38.25}       & 68.91                    & 4.94                     & 5.08                     \\ \bottomrule
\multicolumn{7}{l}{\footnotesize \textbf{Bold} indicates the best result, and \underline{underline} denotes the second- and third-best results.}
\end{tabular}
}
\end{table}

Beyond existing empirical studies on VD that focus solely on evaluating LLM performance under prompt engineering settings, this section emphasizes the performance gains of different post-training methods relative to base models and the gap between VD-specialized LLMs and ultra-large scale LLMs, providing insights for future work on post-training LLMs for VD. Specifically, in addition to post-training methods introduced in Section~\ref{sec:filter}, our evaluation includes representative LLMs from the DeepSeek, Qwen, GPT, and Gemini families, as well as recent VD-specific post-training methods such as R2VUL~\citep{58}, ReVD~\citep{57}, VulnLLM-R~\citep{55}, and MARCO~\citep{60}.

The results in Table~\ref{tab:performance_comparison} demonstrate that under the specification-based reward guidance, the on-policy RL algorithm GRPO outperforms all other post-training baselines. Specifically, Qwen3-4B-OpenVul achieves a 183\% relative improvement in P-Pass@1 over the original Qwen3-4B model, even surpassing Qwen3-235B-A22B and gemini-2.5-flash, which are dozens of times larger. Furthermore, Qwen3-4B-OpenVul’s P-Pass@8 of 47.01\% exceeds all other models, highlighting that on-policy RL, by encouraging self-exploration on difficult samples, effectively improves the accuracy of rollouts sampled during test-time scaling, underscoring its indispensable role in post-training pipelines for LLM-based VD.

Conversely, several VD-specific post-training methods yield suboptimal results. Models like R2VUL and ReVD even underperform relative to the original Qwen3-4B. This decline is attributed to their use of rationalization during SFT data curation (as discussed in Section~\ref{rq:sft_data_curation}), which induces hallucinations. In contrast, DPO and ORPO leverage rejection sampling to ensure higher data integrity, achieving top-tier performance among off-policy optimization methods. Notably, while VulnLLM-R incorporates data filtering, its reliance on CWE types may limit SFT data quality. Similarly, although MARCO trains the policy model using GRPO, its coarse-grained detection-based reward misleads the RL training process, causing it to fall below the SFT baseline. These findings reaffirm that beyond the algorithm, data curation and reward system design are the primary determinants of post-training success in VD.

Finally, while Qwen3-4B-OpenVul still trails behind DeepSeek-V3.1, its performance is remarkably close to its teacher model, DeepSeek-R1-0528. This suggests that future work focusing on higher-quality data distillation could further bridge the gap between small, specialized VD LLMs and general-purpose LLM giants.

\begin{tcolorbox}[colback=orange!10, colframe=orange!100, boxrule=1pt, left=1mm, right=1mm, top=1mm, bottom=1mm]
\textbf{Answer-6:} On-policy RL outperforms all other post-training baselines, with Qwen3-4B-OpenVul significantly improving performance over its base model and larger-scale LLMs. While DPO and ORPO show benefits, current specialized  VD LLMs underperform due to rationalization-induced hallucinations or coarse-grained rewards that mislead RL training, indicating that data curation and fine-grained reward design are the primary determinants of post-training success in VD.
\end{tcolorbox}

\subsection{RQ7: To what extent do different judge LLMs impact the evaluation accuracy?}\label{rq:human}

\begin{wraptable}{r}{0.6\columnwidth}
\caption{Evaluation Bias of Different Judge LLMs}
\label{tab:judge_bias}
\centering
\resizebox{0.58\columnwidth}{!}{
\renewcommand{\arraystretch}{1.0}
\begin{tabular}{ccc}
\toprule
\textbf{Evaluation LLM-Judge} & \textbf{\#Correct Judgments} & \textbf{\# Incorrect Judgments} \\ \midrule
GPT-4.1 mini        & 341                      & 59                          \\
GPT-OSS-20B         & 395                      & 5                           \\
GPT-OSS-120B        & \textbf{399}                      & \textbf{1}                           \\ \bottomrule
\end{tabular}

}
\end{wraptable}
During the preliminary stages of our experiments, we benchmark the evaluative accuracy of three candidate judge LLMs: GPT-4.1 mini, GPT-OSS-20B, and GPT-OSS-120B. For all judge LLMs, the reasoning effort is set to the highest level where applicable. To quantify the potential bias or error introduced by these automated judges, we randomly sample 400 completions from Qwen3-4B and manually verify the LLM judgments. The results in Table~\ref{tab:judge_bias} reveal that for the 400 vulnerable samples, GPT-4.1 mini incorrectly labeled 59 model predictions as correct (a 14.75\% error rate). Our analysis shows that most of these failures occur when the model's predicted vulnerability type is similar to or identical to the ground truth, but the underlying reasoning is flawed. In these instances, GPT-4.1 mini fails to rigorously audit the model's root cause analysis, leading to the biased judgement.

In contrast, GPT-OSS-20B and GPT-OSS-120B achieve a judgment accuracy of over 99\% across these 400 samples. The occasional errors stem from hallucinations regarding the context. For instance, the judge LLM claims that the model fails to detect any vulnerabilities even when the model has explicitly predicted a vulnerability. Furthermore, we note that GPT-OSS-20B frequently produces repetitive content in certain scenarios. Consequently, we select GPT-OSS-120B as the default judge LLM for this work. This choice ensures the stability and accuracy of the reward assessment during RL training and the performance evaluation during model testing.

\begin{wraptable}{r}{0.6\columnwidth}
\caption{Performance Evaluation by an Independent Judge}
\label{tab:judge_independence}
\centering
\resizebox{0.58\columnwidth}{!}{
\renewcommand{\arraystretch}{1.0}
\begin{tabular}{ccccc}
\toprule
Evaluation LLM-Judge & P-Pass@1 & P-Pass@8 & F1 & Cohen's Kappa \\ \midrule
GPT-OSS-120B & 21.08 & 47.01 & 38.25 & \multirow{2}{*}{0.8537} \\
Qwen3.5-122B-A10B & 20.89 & 48.63 & 38.45 &  \\ \bottomrule
\end{tabular}
}
\end{wraptable}

\noindent\textbf{Mitigating Judge Overfitting via Independent Evaluation.} To address concerns of potential judge overfitting, where policy model (i.e., Qwen3-4B-OpenVul evaluated in Table~\ref{tab:performance_comparison}) may achieve inflated gains by merely exploiting the reward judge's specific preferences, we decouple the reward and evaluation judges by utilizing GPT-OSS-120B exclusively for reward signaling during RL and Qwen3.5-122B-A10B as an independent judge of similar scale for the final performance assessment. As shown in Table~\ref{tab:judge_independence}, the high Cohen's Kappa of 0.8537 between the two judge LLMs proves that the observed performance gains are robust and not a result of judge overfitting.

\begin{tcolorbox}[colback=orange!10, colframe=orange!100, boxrule=1pt, left=1mm, right=1mm, top=1mm, bottom=1mm]
\textbf{Answer-7:} GPT-OSS-120B offers a superior balance of evaluative precision and output stability for high-fidelity VD research. Moreover, the strong inter-judge agreement after decoupling the reward and evaluation judges confirms that the performance gains are robust and not a result of judge overfitting.
\end{tcolorbox}

\subsection{RQ8: How does model assessment shift across multiple evaluation granularities?}\label{rq:sankey}

\begin{wrapfigure}{r}{0.55\columnwidth}
\centering
\includegraphics[width=0.52\columnwidth]{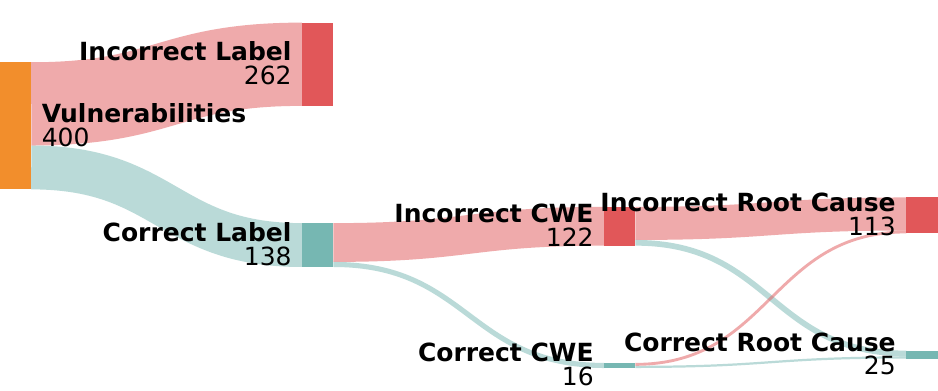}
\caption{Model Assessment Shifts Across Multi-Level Evaluation Granularities} \label{fig:pov_sankey}
\vspace{-0.2in}
\end{wrapfigure}

In this subsection, we conduct a cross-granularity analysis of 400 randomly selected Qwen3-4B completions to investigate how model assessment shifts as evaluation criteria are refined. We examine three granularities: detection-level binary comparison, prediction-level CWE match, and reasoning-level root-cause judgment.

As illustrated in Figure~\ref{fig:pov_sankey}, while the model correctly identified 138 vulnerabilities at the binary-label level, 88.4\% of these are mislabeled at the CWE level, indicating that detection-level binary comparison significantly overestimate performance. Furthermore, among the 16 correct CWE predictions, 56.3\% relies on the reasoning-level root-cause analyses unrelated to the ground truth. Conversely, our analysis reveals that prediction-level evaluation is often biased: 6.6\% of predictions with mismatched CWE types actually identify the correct root cause, suggesting that ambiguous CWE matching is an imperfect proxy for reliable evaluation. These findings uncover that detection- and prediction-level evaluations systematically inflate or bias reported performance. Accurate VD assessment necessitates reasoning-level evaluation to confirm that the model genuinely comprehends the vulnerability. This level of granularity is also essential for RL training to prevent models from being rewarded for flawed reasoning that guesses a correct label.

\begin{tcolorbox}[colback=orange!10, colframe=orange!100, boxrule=1pt, left=1mm, right=1mm, top=1mm, bottom=1mm]
\textbf{Answer-8:} Relying on binary comparison and CWE match significantly overestimate model performance. True model capability is only captured at the reasoning level, which reveals both lucky guesses and cases where the model correctly identifies root causes despite failing under strict CWE matching.
\end{tcolorbox}

\section{Related Work}
\subsection{Prompt-based LLMs for VD}

Recent advancements in LLM-based VD have evolved from single-agent structure-aware prompting methods, which incorporate code dependencies and semantic reasoning to address the limitations of linear text processing~\citep{1,3,4,6}, to multi-agent frameworks designed to mitigate hallucinations and overconfidence. Techniques such as adaptive prompting~\citep{2} and CoT reasoning~\citep{7,30} improve root-cause analysis by guiding models through explicit reasoning steps, yet extensive evaluations show that LLMs still struggle with out-of-distribution generalization and complex vulnerability reasoning, with performance varying notably between zero-shot prompting and domain-specific fine-tuning~\citep{16,17,18,19,20,21,22,23,24,25,26,27,28,29,30,31}. To address these limitations, multi-agent systems introduce collaborative and adversarial mechanisms, such as role-specific agents, mock-court debates~\citep{8}, sequential decision-making~\citep{13}, and interactive refinement loops~\citep{10,11}, enabling cross-verification for improved precision~\citep{8,9,10,11,12,13}. However, they remain susceptible to noise accumulation in multi-turn interactions, which can distract from core reasoning.

\subsection{Post-training LLMs for VD}

\textbf{SFT.} Early efforts primarily relied on SFT with task- or instruction-augmented data. VulLLM~\citep{53} jointly trained vulnerability detection, classification, and explanation tasks under unified instruction templates, which improved generalization to unseen projects and vulnerability types. MSIVD~\citep{54} further reduced annotation costs by automatically synthesizing diverse vulnerability-related instructions and responses. VulnLLM-R~\citep{55} incorporated structured reasoning supervision together with an agent scaffold that decomposed vulnerability analysis into multi-stage reasoning pipelines.

\noindent\textbf{Off-policy Preference Optimization.} To overcome the limitations of SFT in distinguishing subtle vulnerabilities, Smart-LLaMA-DPO~\citep{56} applied DPO using constructed preference pairs to align models toward correct and explainable vulnerability analyses, particularly in the smart contract domain. ReVD~\citep{57} integrated curriculum preference optimization into DPO by constructing preference pairs from synthetic reasoning data. R2Vul~\citep{58} combined structured reasoning distillation from teacher models with ORPO. However, these methods generally relied on static preference datasets, restricting the model's exploration of diverse reasoning paths.

\noindent\textbf{On-policy RL.} More recently, RL was explored for VD. RLFD~\citep{59} formulated VD using a conventional RL setup on relatively small pre-trained language models to improve fine-grained detection and localization. Another study~\citep{60} applied GRPO to post-train LLMs, where the RL training relied primarily on detection-level label correctness as the reward signal.
\section{Discussion and Future Work}

\noindent\textbf{Evaluation Scope}. Due to the high computational cost of post-training, where a single RL run can take several days, we focus on a representative set of methods and limit our experiments to the Qwen3-4B backbone and C/C++ data. Accordingly, some generic method extensions (e.g., SimPO~\citep{88}, Dr.GRPO~\citep{85}) and hyperparameters (e.g., the number of generations in GRPO~\citep{83}) that are not directly relevant to VD objectives are omitted. Constructing cross-language VD datasets with full repository-level context also remains challenging. Nevertheless, our methodology is model-agnostic and language-independent, and we open-source our framework to support future validation.

\noindent\textbf{Evaluation Validity.} While potential data leakage between pre-training and evaluation may bias absolute performance, our goal is to demonstrate relative improvements. Results consistently show that post-training significantly improves VD performance over the base model. Given the scarcity of recent vulnerability data and the rapid evolution of LLMs, developing more robust evaluation protocols remains an important direction for future work.
\section{Conclusion}
This paper presented a systematic study of post-training pipeline for LLM-based VD, covering cold-start SFT, off-policy preference optimization, and on-policy RL. Our findings revealed that while SFT establishes foundational capabilities, the choice of data curation, specifically rejection sampling over rationalization, is critical to prevent reasoning hallucinations. We demonstrated that heavy SFT benefits off-policy preference optimization but can inhibit the self-exploration necessary for on-policy GRPO. Furthermore, we showed that fine-grained, root-cause-aligned rewards significantly outperform coarse binary-level and CWE-level signals in training efficiency and accuracy. Our results indicated that on-policy RL enables models to surpass SFT-based and even SOTA zero-shot LLMs. Finally, we highlighted necessity of reasoning-level evaluation protocols to mitigate performance overestimation. By open-sourcing our framework and datasets, we provided a standardized foundation and practical guidelines for developing specialized VD LLMs.

\bibliography{main}
\clearpage
\appendix
\section{Details of Dataset}\label{sec:datasets}

\begin{figure}[h]\centering
\includegraphics[width=0.75\columnwidth]{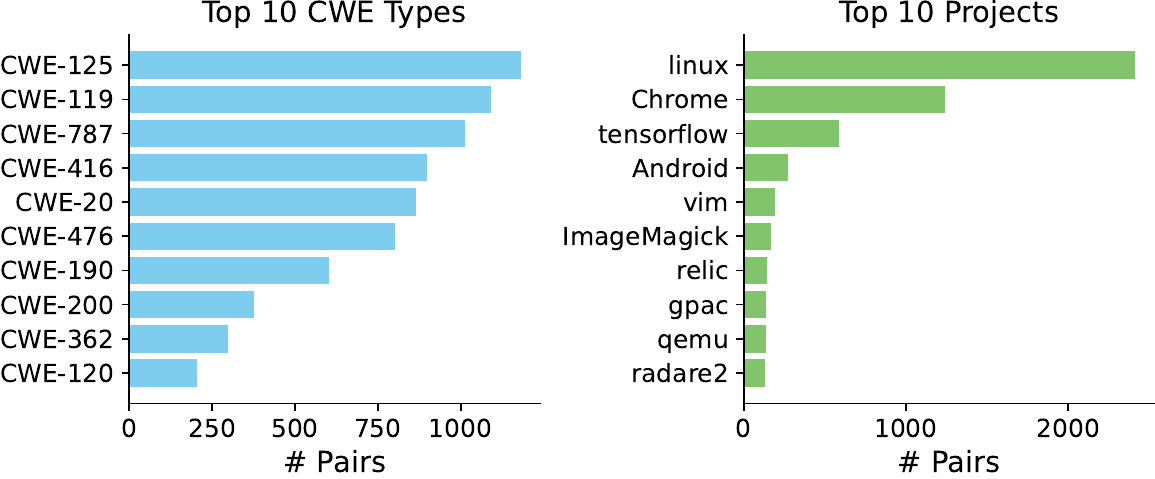}
\caption{Distribution of Dataset} \label{fig:cwe}
\end{figure}

Figure~\ref{fig:cwe} shows the distribution of the top 10 vulnerability types and the top 10 projects in our dataset.

\section{Details of CWE Matching}\label{sec:cwe_match}

\begin{figure}[h]\centering
\includegraphics[width=0.65\columnwidth]{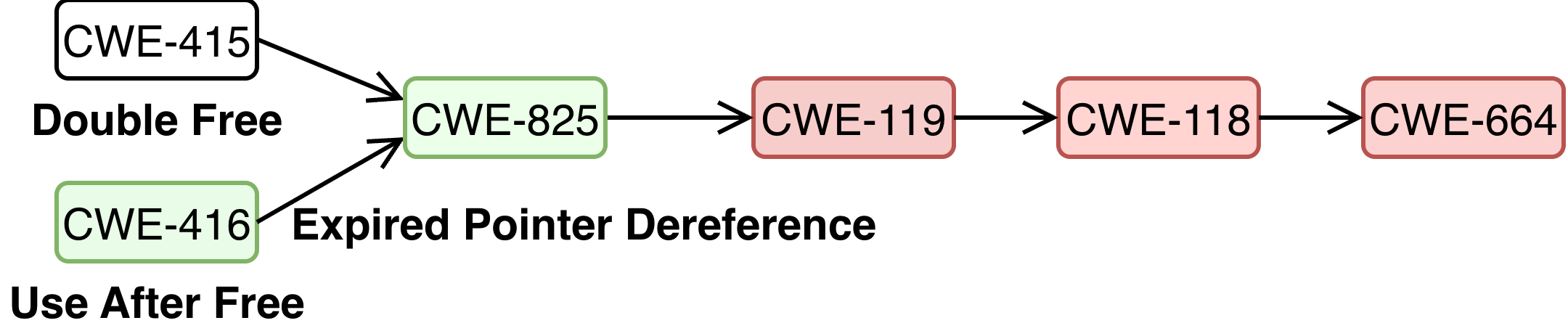}
\caption{Example of Prediction-Level Evaluation Logic for CWE-416 within the CWE-1000 Taxonomy} \label{fig:CWE}
\end{figure}

For prediction-based rewarding in RL training and prediction-level matching in model evaluation, we avoid an exact CWE-match approach due to the hierarchical nature of the CWE. Following the mapping methodology in~\citep{61,62}, a prediction is deemed correct if the predicted CWE ID and the ground truth share a direct parent-child relationship within the CWE Research Concepts view (CWE-1000) taxonomy~\citep{87}. For example, Figure~\ref{fig:CWE} illustrates that CWE-415 and CWE-416 share a direct parent node, CWE-825. When the ground truth is CWE-416, the prediction is considered correct as long as the model's predicted vulnerability types includes CWE-416 or CWE-825. However, if the model predicts only higher-level nodes such as CWE-119, CWE-118, or CWE-664, the prediction is treated as incorrect.
\section{Implemetation Details of LLM Prompts}\label{sec:appendix_judge}

Figure~\ref{box:vul_input}-\ref{box:sample_judge_fix} presents the prompts implemented for VD, SFT data curation, reasoning- and specification-level LLM judgments, and sample specification generation. They are slightly modified for better presentation.

\subsection{Prompt for LLM-based VD}
\begin{figure}[H]
\begin{tcolorbox}[boxrule=1pt, left=1mm, right=1mm, top=1mm, bottom=1mm, fontupper=\small]
<|system|> You are a vulnerability detection expert specializing in identifying security flaws in C/C++ code, with a focus on Common Weakness Enumeration (CWE) standards. You provide precise, evidence-based analysis without speculation, and clearly label any vulnerabilities you detect.\\
<|user|>\\
Your task is to evaluate whether the following C/C++ code contains any security vulnerabilities.\\\\
You will be provided with two sections:\\
1. Context: Relevant code such as includes, type definitions, global variables, macros, and definitions of any functions called within the target function.\\
2. Code: The target function to analyze.\\\\
Use all available information to analyze the function step by step.\\
If the target function alone is insufficient to determine whether a vulnerability exists, refer to the Context section before making a judgment.\\
Do not assume vulnerabilities — only report what is supported by the code and context.\\\\
In your final response, list all detected vulnerabilities and CWE identifiers if applicable.\\
Conclude with one of the following indicators on a new line:\\
- \texttt{HAS\_VUL} — if any vulnerabilities are found\\
- \texttt{NO\_VUL} — if no vulnerabilities are found\\\\
\verb|```|Context\\
\{Includes\}\\
\{Type Definitions\}\\
\{Called Methods\}\\
\verb|```|\\\\
\verb|```|Code\\
File: ...\\
Method: ...\\
----------------------------------------\\
\{Target Function\}\\
\verb|```|\\
Analyze the code now.
\end{tcolorbox}
\caption{Prompt for LLM-based VD}
\label{box:vul_input}
\end{figure}

Figure~\ref{box:vul_input} illustrates the workflow of VD. The LLM detector is instructed to detect vulnerabilities in C/C++ code. This involves analyzing a target function alongside its provided context to identify flaws like memory leaks, buffer overflows, or logic errors. The process requires a step-by-step evaluation of how data flows through the code, ensuring that every identified vulnerability is supported by evidence and mapped to a specific CWE identifier. The final output lists all detected vulnerabilities and concludes with \texttt{HAS\_VUL} if any are found, or \texttt{NO\_VUL} if no vulnerabilities are found.

\subsection{Prompt for SFT Data Curation}

\begin{figure}[H]
\begin{tcolorbox}[boxrule=1pt, left=1mm, right=1mm, top=1mm, bottom=1mm, fontupper=\scriptsize]
<|system|> You are a vulnerability detection expert specializing in identifying security flaws in C/C++ code, with a focus on Common Weakness Enumeration (CWE) standards. You provide precise, evidence-based analysis without speculation, and clearly label any vulnerabilities you detect.\\
<|user|>\\
\textbf{1. Task}\\
Your task is to generate a Vulnerability Reasoning Trace for a specific code snippet. This reasoning trace will be used to train a student model to detect vulnerabilities.\\

Crucially, you must simulate a "Blind Audit". You are provided with Ground Truth information (Code status, CWE ID, CVE description, commit message, patch diff) to ensure your analysis is factually correct, but your output must appear as if you derived the conclusion solely through deductive code analysis.\\

\textbf{2. Input Data}\\
\{INPUT\}\\

\verb|```|Hidden Ground Truth CONFIDENTIAL - FOR TEACHER CONTEXT ONLY The following information is the "Answer Key". Use it to verify which lines are vulnerable and why, but NEVER reference these documents, IDs, or the existence of a patch in your final output.\\

- Code Status: \{CODE\_STATUS\}\\
- CWE ID: \{CWE\_ID\}\\
- CVE Description: \{CVE\_DESCRIPTION\}\\
- Commit Message: \{COMMIT\_MESSAGE\}\\
- Patch Diff: \{CODE\_DIFF\}\\
\verb|```|\\

\textbf{3. Analysis Instructions}\\
Step 1: Analyze the Target Code: Examine the Code and its Context.\\
Step 2: Check Code Status: Look at the Code Status provided above.\\
- If PRE-PATCH: Your reasoning must explain how the code enables the specific vulnerability described in the Ground Truth.\\
- If POST-PATCH: Your reasoning must explain why the code is safe, specifically highlighting the presence of the validation or logic that prevents the vulnerability described in the Ground Truth.\\
Step 3: Consult Ground Truth (Silently): Read the Hidden Ground Truth to understand the vulnerability mechanics. Use this strictly as a fact-checking reference.\\
Step 4: Simulate Discovery: Construct a step-by-step logical argument that leads to that finding using only the visible code tokens.\\
- Do not say: "The patch fixes this by..."\\
- Do say: "The variable len is validated against MAX\_SIZE before use, preventing overflow..." (if Fixed) or "The variable len is used without validation..." (if Vulnerable).\\
Step 5:  Determine Verdict:\\
- If `Code Status` is "PRE-PATCH": Verdict is \texttt{HAS\_VUL}.\\
- If `Code Status` is "POST-PATCH": Verdict is \texttt{NO\_VUL}.\\
        
\textbf{4. Negative Constraints (Strict Adherence Required)}\\
- NO mentions of Hidden Ground Truth, such as "CVE", "Commit", "Patch", "Diff", "Fix", or "Description".\\
- NO references to "The provided info" or "Ground truth".\\
- NO phrasing like "As seen in the diff" or "This was later patched".\\
- NO external knowledge hallucination (e.g., do not invent a specific exploit date or hacker group). Stick to the code logic.\\
    
\textbf{5. Output Format}\\
In your final response, list all detected vulnerabilities and CWE identifiers if applicable.\\
Conclude with one of the following verdicts on a new line:\\
- \texttt{HAS\_VUL} — if any vulnerabilities are found\\
- \texttt{NO\_VUL} — if no vulnerabilities are found
\end{tcolorbox}
\caption{Prompt for Rationalization-based Data Curation}
\label{box:rationalization}
\end{figure}

\subsubsection{Rejection Sampling}
Since the teacher LLM's input in the rejection sampling-based data curation does not include ground truth vulnerability information, its prompt is identical to the one used for LLM-based VD shown in Figure~\ref{box:vul_input}.

\subsubsection{Rationalization}

Figure~\ref{box:rationalization} presents the prompt designed for rationalization-based data curation, specifically for generating high-quality vulnerability reasoning CoTs to train specialized VD LLMs. The strategy employs a teacher-student framework where a powerful LLM acts as the teacher. While the teacher model is provided with \texttt{Hidden Ground Truth} (CVE details, commit messages, and patch diffs) to ensure factual accuracy, it is strictly instructed to simulate a blind audit. This means the teacher model must reverse-engineer a logical justification for the vulnerability (or its absence) using only the code's internal logic, without explicitly mentioning the external metadata.

\subsection{Judge Prompts for Reasoning-based Rewards and Model Evaluation}\label{sec:reasoning}
\subsubsection{Vulnerable Version}

The prompt in Figure~\ref{box:judge_vul} for our judge LLM provides a structured framework for evaluating detection LLM's analysis of vulnerable code. The input consists of a JSON object containing the model-generated analysis and a ground truth reference, including the vulnerability status, official CVE description, and patch details. The evaluation is organized according to a reasoning-level correctness rubric.

Specifically, the analysis is judged as \texttt{CORRECT} only if it both identifies the code as vulnerable and correctly explains the root cause of the vulnerability. \texttt{PARTIALLY INCORRECT} indicates that the analysis recognizes the code as vulnerable but misidentifies the root cause; this distinction is used to differentiate quality, but it is treated as an incorrect prediction during RL rewarding and performance evaluation. Finally, the output required from the LLM is a JSON object that provides a brief justification and selects the corresponding rubric option.

\subsubsection{Patched Version}

The prompt in Figure~\ref{box:judge_fix} is similar to the previously described evaluation framework for vulnerable code samples but is adapted for post-patched code where the target CVE has already been fixed. The input consists of a JSON object containing a model-generated analysis and ground truth information, including the target CVE's absence in the code, its description, the patch commit message, and the patch diff.

The evaluation is based on the reasoning-level correctness rubric. Notably, the distinction between \texttt{CORRECT} and \texttt{UNKNOWN} is used to differentiate whether the analysis correctly identifies the absence of the original CVE or only finds other unrelated vulnerabilities. \texttt{UNKNOWN} does not indicate a correct identification of the target CVE, but it is separated to reflect discovery of different vulnerabilities. Importantly, \texttt{UNKNOWN} is treated as acceptable because, based on the ground truth, we can only confirm that the target CVE has been fixed, but we cannot determine whether other unknown vulnerabilities exist; therefore, the LLM may validly report other potential issues without being considered incorrect. Finally, the LLM is required to produce a structured JSON object containing the justification and the selected option.

\begin{figure}[H]
\begin{tcolorbox}[boxrule=1pt, left=1mm, right=1mm, top=1mm, bottom=1mm, fontupper=\scriptsize]
<|system|> You are to act as a meticulous and impartial Code Security Expert and Evaluator.\\
<|user|>\\
\textbf{1. Goal}\\
Your primary goal is to assess the quality of an analysis of a vulnerable piece of code. You must evaluate the analysis against a provided set of ground truth information. Your judgment must be objective, strictly adhering to the provided option rubric and based only on the information given.\\\\
\textbf{2. Input Format}\\
You will be provided with a JSON object containing two main keys: analysis and ground\_truth\_info\\
\verb|```|json\\
\{\\
\hspace*{1em}"analysis": "<The full analysis, including its reasoning and answer.>",\\
\hspace*{1em}"ground\_truth\_info": \{\\
\hspace*{2em}"is\_vulnerable": true,\\
\hspace*{2em}"cve\_description": "<The official CVE description of the vulnerability.>",\\
\hspace*{2em}"patch\_commit\_message": "<The developer's commit message that may explain the vulnerability.>"\\
\hspace*{2em}"patch\_commit\_diff": "<A git-style diff showing the changes from the pre-patched (vulnerable) to the post-patched (non-vulnerable) code.>"\\
\hspace*{1em}\}\\
\}\\
\verb|```|\\\\
Analysis Context \& Critical Warning:\\
The analysis you are evaluating is generated based on the pre-patched (vulnerable) code. The patch\_commit\_diff and patch\_commit\_message is provided only as a reference to help you understand the precise location and nature of the ground truth vulnerability. Do not let it mislead you into thinking the vulnerability has already been fixed in the pre-patched code that is analyzed.\\\\
\textbf{3. Evaluation Workflow and Option Rubric}\\
You must follow these steps to evaluate the analysis and produce a final JSON output. For each dimension, you need to provide a brief justification and choose an option.\\
Step 1: Analyze Ground Truth\\
First, carefully review all the information in the ground\_truth\_info. This is your foundation for judgment.\\
Step 2: Evaluate Each Dimension\\\\
Assess the analysis across the following dimension. Choose an option for each based on the rubric below.\\
Dimension 1: Correctness\\
Task: Evaluate if the analysis correctly identifies the target CVE mentioned in the ground\_truth\_info.\\
Option Rubric:\\
* \texttt{CORRECT}: The analysis identifies the code as vulnerable, AND the explanation of the root cause of the predicted vulnerability also aligns with the ground truth vulnerability information provided in the ground\_truth\_info. Besides, it is acceptable if the analysis also identifies other vulnerabilities.\\
* \texttt{PARTIALLY INCORRECT}: The analysis identifies the code as vulnerable, BUT the explanation of the root cause of the predicted vulnerability does not align with the ground truth vulnerability information provided in the ground\_truth\_info.\\
* \texttt{INCORRECT}: The analysis identifies the code as non-vulnerable.\\\\
\textbf{4. Output Format}\\
Your final output must be a single JSON object. Do not include any text or explanation outside of the JSON structure. The JSON must contain a key for each dimension's justification and option.\\
\verb|```|json\\
\{\\
\hspace*{1em}"correctness": \{\\
\hspace*{2em}"justification": "<Your brief reason>",\\
\hspace*{2em}"option": <choose from ["\texttt{CORRECT}", "\texttt{PARTIALLY INCORRECT}", "\texttt{INCORRECT}"]>\\
\hspace*{1em}\}\\
\}\\
\verb|```|\\\\
\textbf{Current Input}\\
\verb|```|json\\
\{INPUT\}\\
\verb|```|
\end{tcolorbox}
\caption{Reasoning-level Judge Prompt for Vulnerable Samples}
\label{box:judge_vul}
\end{figure}

\begin{figure}[H]
\begin{tcolorbox}[boxrule=1pt, left=1mm, right=1mm, top=1mm, bottom=1mm, fontupper=\scriptsize]
<|system|> You are to act as a meticulous and impartial Code Security Expert and Evaluator.\\
<|user|>\\
\textbf{1. Goal}\\
Your primary goal is to assess the quality of an analysis of post-patched code in which the target CVE has been fixed. You must evaluate the analysis against a provided set of ground truth information. Your judgment must be objective, strictly adhering to the provided option rubric and based only on the information given.\\

\textbf{2. Input Format}\\
You will be provided with a JSON object containing two main keys: analysis and ground\_truth\_info\\
\verb|```|json\\
\{\\
\hspace*{1em}"analysis": "<The full analysis, including its reasoning and answer.>",\\
\hspace*{1em}"ground\_truth\_info": \{\\
\hspace*{2em}"target\_CVE\_in\_code": false,\\
\hspace*{2em}"cve\_description": "<The official CVE description of the vulnerability that was fixed.>",\\
\hspace*{2em}"patch\_commit\_message": "<The developer's commit message that may explain the fix.>",\\
\hspace*{2em}"patch\_commit\_diff": "<A git-style diff showing the changes that fixed the vulnerability.>"\\
\hspace*{1em}\}\\
\}\\
\verb|```|\\

Analysis Context \& Critical Warning:\\
The analysis you are evaluating is generated based on the post-patched code in which the target CVE has been fixed. The ground\_truth\_info is provided only as a reference to help you understand how the target CVE is fixed. Do not let it mislead you into thinking the target CVE is still present in the post-patched code that is analyzed. Please note that "target\_CVE\_in\_code": false in the ground\_truth\_info can only indicate that the target CVE does not exist in the code, but it cannot guarantee whether the code contains other unknown vulnerabilities.\\

\textbf{3. Evaluation Workflow and Option Rubric}\\
You must follow these steps to evaluate the analysis and produce a final JSON output. For each dimension, you need to provide a brief justification and choose an option.\\
Step 1: Analyze Ground Truth\\
First, carefully review all the information in the ground\_truth\_info. This is your foundation for judgment.\\
Step 2: Evaluate Each Dimension\\

Assess the analysis across the following four dimensions. Choose an option for each based on the rubric below.\\
Dimension 1: Correctness\\
Task: Evaluate whether the analysis identifies that a vulnerability with the exactly same root cause in ground\_truth\_info still exists in the post-patched code.\\
Option Rubric:\\
- \texttt{CORRECT}: The analysis finds no vulnerabilities in the code.\\
- \texttt{UNKNOWN}: The analysis does not identify a vulnerability with the exactly same root cause as in ground\_truth\_info, but it does identify other unknown vulnerabilities with different root causes.\\
- \texttt{INCORRECT}: Select this option ONLY IF the analysis identifies that a vulnerability with the exactly same root cause as in ground\_truth\_info still exists in the code.\\
Please select \texttt{UNKNOWN} if the analysis identifies vulnerabilities whose root causes are not exactly the same as the vulnerability in ground\_truth\_info, even if they are only similar. For example, the analysis identifies that the code contains an out-of-bound access vulnerability, and the target CVE in ground\_truth\_info is also an out-of-bound access vulnerability. However, the root causes of the two vulnerabilities are not exactly same (e.g., they occur in different locations). In this situation, you should choose \texttt{UNKNOWN}, because the two vulnerabilities are not exactly the same.\\

\textbf{4. Output Format}\\
Your final output must be a single JSON object. Do not include any text or explanation outside of the JSON structure. The JSON must contain a key for each dimension's justification and option.\\
\verb|```|json\\
\{\\
\hspace*{1em}"correctness": \{\\
\hspace*{2em}"justification": "<Your brief reason>",\\
\hspace*{2em}"option": <choose from ["\texttt{CORRECT}", "\texttt{UNKNOWN}", "\texttt{INCORRECT}"]>\\
\hspace*{1em}\}\\
\}\\
\verb|```|\\\\
\textbf{Current Input}\\
\verb|```|json\\
\{INPUT\}\\
\verb|```|
\end{tcolorbox}
\caption{Reasoning-level Judge Prompt for Patched Samples}
\label{box:judge_fix}
\end{figure}

\subsection{Prompts for Sample Specification Generation}\label{sec:generation}

\begin{figure}[H]
\begin{tcolorbox}[boxrule=1pt, left=1mm, right=1mm, top=1mm, bottom=1mm, fontupper=\footnotesize]
\textbf{1. Context}\\
We have a code sample with a known vulnerability. We need to check if a detector successfully "recalls" (finds) this specific issue.\\

\textbf{2. Input Data}\\
- Target Function and Its Contexts:\{INPUT\}\\
- Code Status: \{CODE\_STATUS\}\\
- CVE Description: \{CVE\_DESCRIPTION\}\\
- Commit Message: \{COMMIT\_MESSAGE\}\\
- Patch Diff: \{CODE\_DIFF\}\\

\textbf{3. Task}\\
Generate a JSON Checklist with these specific dimensions:\\

Dimension 1: Verdict\_Recall\\
- Create a check: "Does the analysis report contain a vulnerability in the target function that matches the semantic type [INSERT\_VULN\_TYPE\_FROM\_CVE]?"\\
- Instruction: Specify that finding this type is sufficient, even if other types are also listed.\\

Dimension 2: Evidence\_Insecure\_Code\\
- Identify the unique code snippet from the Diff (lines marked \verb|`|-\verb|`|) that causes the bug.\\
- Create a check: "Does the analysis explicitly cite the code snippet [INSERT\_UNIQUE\_CODE\_SNIPPET] as the source of the issue?"\\

Dimension 3 : Reasoning\_Mechanism\\
- Create a check: "Does the analysis explain the trigger condition? Specifically, that [INSERT\_SPECIFIC\_TRIGGER] (e.g., 'input length exceeds buffer' or 'integer wraps around') leads to the vulnerability."\\

\textbf{4. Output Format (JSON Only)}\\
\verb|```|json\\
\{\\
\hspace*{1em}"phase": "pre\_patch",\\
\hspace*{1em}"checklist": [\\
\hspace*{2em}\{\\
\hspace*{3em}"dimension": "Verdict\_Recall",\\
\hspace*{3em}"description": "The analysis report includes a finding for the target function that semantically matches [INSERT\_EXTRACTED\_TYPE] (e.g., Integer Overflow)."\\
\hspace*{2em}\},\\
\hspace*{2em}\{\\
\hspace*{3em}"dimension": "Evidence\_Insecure\_Code",\\
\hspace*{3em}"description": "The analysis explicitly cites the code snippet [INSERT\_UNIQUE\_CODE\_SNIPPET] (e.g., 'memcpy(dest, src, len)') as problematic."\\
\hspace*{2em}\},\\
\hspace*{2em}\{\\
\hspace*{3em}"dimension": "Reasoning\_Mechanism",\\
\hspace*{3em}"description": "The analysis explains that the vulnerability exists because [INSERT\_BRIEF\_LOGIC] (e.g., length check is missing)."\\
\hspace*{2em}\}\\
\hspace*{1em}]\\
\}\\
\verb|```|
\end{tcolorbox}
\caption{Specification Generation Prompt for Vulnerable Samples}
\label{box:rubric_vul}
\end{figure}
\begin{figure}[H]
\begin{tcolorbox}[boxrule=1pt, left=1mm, right=1mm, top=1mm, bottom=1mm, fontupper=\footnotesize]
\textbf{1. Context}\\
The code is \texttt{NO\_VUL} regarding a specific previous vulnerability (Fixed).
We need to verify if the detector recognizes this safety or at least does not hallucinate the old bug.\\

\textbf{2. Input Data}\\
- Target Function and Its Contexts:\{INPUT\}\\
- Code Status: \{CODE\_STATUS\}\\
- CVE Descriptio: \{CVE\_DESCRIPTION\}\\
- Commit Message: \{COMMIT\_MESSAGE\}\\
- Patch Diff: \{CODE\_DIFF\}\\

\textbf{3. Task}\\
Generate a JSON Checklist with these specific dimensions:\\
Dimension 1: Verdict\_Absence\_of\_Specific\_Vuln\\
- Reflect the logic: "The analysis is correct if it concludes NO\_VUL OR if the detected vulnerabilities DO NOT include [INSERT\_OLD\_VULN\_TYPE]."\\

Dimension 2: Evidence\_Safeguard\_Code\\
- Identify the new check/sanitizer from the Diff (lines marked \verb|`|+\verb|`|).\\
- Create a check: "Does the analysis cite the code snippet [INSERT\_SAFEGUARD\_SNIPPET] as a safety factor?"\\

Dimension 3: Reasoning\_Resolution\\
- Create a check: "Does the analysis explain that the code is safe because [INSERT\_FIX\_LOGIC]?"\\

\textbf{4. Output Format (JSON Only)}\\
\verb|```|json\\
\{\\
\hspace*{1em}"phase": "post\_patch",\\
\hspace*{1em}"checklist": [\\
\hspace*{2em}\{\\
\hspace*{3em}"dimension": "Verdict\_Absence\_of\_Specific\_Vuln",\\
\hspace*{3em}"description": "The analysis either concludes the function is NO\_VUL, OR its list of detected vulnerabilities does NOT include '[INSERT\_OLD\_VULN\_TYPE]'."\\
\hspace*{2em}\},\\
\hspace*{2em}\{\\
\hspace*{3em}"dimension": "Evidence\_Safeguard\_Code",\\
\hspace*{3em}"description": "The analysis explicitly cites the code snippet [INSERT\_SAFEGUARD\_SNIPPET] (e.g., 'if (size > MAX)') as a mitigating factor."\\
\hspace*{2em}\},\\
\hspace*{2em}\{\\
\hspace*{3em}"dimension": "Reasoning\_Resolution",\\
\hspace*{3em}"description": "The analysis explains that the potential issue is prevented by [INSERT\_LOGIC\_SUMMARY] (e.g., validating input size)."\\
\hspace*{2em}\}\\
\hspace*{1em}]\\
\}
\verb|```|
\end{tcolorbox}
\caption{Specification Generation Prompt for Patched Samples}
\label{box:rubric_fix}
\end{figure}

\subsubsection{Vulnerable Version}\label{sec:generation_vul}
Figure~\ref{box:rubric_vul} presents the prompt designed to guide an LLM-as-a-Judge in evaluating the quality of a vulnerability detector's analysis on pre-patch vulnerable code. It instructs an LLM to generate a structured JSON checklist based on the provided vulnerable function, CVE description, commit message, and patch diff. The checklist focuses on three critical validation dimensions: \texttt{Verdict Recall} (verifying the detector identifies the correct vulnerability type), \texttt{Evidence Identification} (ensuring the detector cites the specific lines of code that were removed or changed in the patch), and \texttt{Reasoning Mechanism} (confirming the detector accurately explains the logic or trigger condition that leads to the exploit).

\subsubsection{Patched Version}\label{sec:generation_fix}

Figure~\ref{box:rubric_fix} presents the prompt designed to guide an LLM-as-a-Judge in evaluating the quality of a vulnerability detector's analysis on patched (vulnerability-fixed) code to ensure it recognizes security improvements and avoids hallucinating resolved bugs. By utilizing the provided post-patch function, fixed-CVE description, commit message, and patch diff, the prompt instructs an LLM to generate a JSON checklist focused on three defensive dimensions: \texttt{Verdict Absence} (verifying the detector does not falsely report the original vulnerability), \texttt{Evidence of Safeguards} (ensuring the detector identifies the specific new lines of code that provide the fix), and \texttt{Reasoning Resolution }(confirming the detector understands how the logic change prevents the exploit).

\subsection{Judge Prompts for Specification-based Rewards}\label{sec:specification}

\subsubsection{Vulnerable Version}
Figure~\ref{box:sample_judge_vul} presents the prompt designed to evaluate a detector's performance on known-vulnerable code by comparing its output against a pre-defined ground truth checklist (introduced in Appendix~\ref{sec:generation_vul}). The evaluation is categorized into three specific dimensions with a nuanced scoring rubric. The \texttt{Verdict} dimension follows a strict binary logic to determine if the specific vulnerability type was identified, regardless of any extraneous findings. The \texttt{Evidence} dimension assesses the precision of the detector's vulnerability localization, distinguishing between exact code snippets, partial matches (such as variable names), and complete misses. Finally, the \texttt{Reasoning} dimension validates the detector's understanding of the attack mechanism, requiring a clear explanation of both the root cause and the technical consequence. The output is formatted as a structured JSON object containing justifications and scores.

\subsubsection{Patched Version}

The prompt presented in Figure~\ref{box:sample_judge_fix} instructs an LLM-as-a-Judge to evaluate a vulnerability detector's performance on patched code, focusing on its ability to recognize security fixes and avoid false positives. The judge LLM compares the detector's candidate analysis against a pre-defined ground truth checklist (introduced in Appendix~\ref{sec:generation_fix}) for safe code across three key metrics. The \texttt{Verdict} dimension uses a binary check to ensure the detector does not hallucinate the specific fixed vulnerability. The \texttt{Evidence dimension} measures how accurately the detector identifies the new safeguard code (e.g., input sanitizers or boundary checks) introduced in the patch. The \texttt{Reasoning} dimension evaluates whether the detector correctly interprets the safety logic, confirming that it understands how the code change successfully mitigates the original threat. The resulting JSON output provides a structured, justifiable assessment of the detector's precision in post-patch scenarios.

\begin{figure}[H]
\begin{tcolorbox}[boxrule=1pt, left=1mm, right=1mm, top=1mm, bottom=1mm, fontupper=\footnotesize]
<|system|> You are to act as a meticulous and impartial Code Security Expert and Evaluator.\\
<|user|>\\
\textbf{1. Goal}\\
Your task is to evaluate a vulnerability analysis for a code sample known to be \texttt{HAS\_VUL} (vulnerable). Verify if the detector successfully identified the specific target vulnerability described in the Checklist.\\

\textbf{2. Scoring Rubric}\\
Dimension 1: Verdict (Strict Binary)\\
- \texttt{CORRECT:} The analysis identifies the specific vulnerability type requested in the checklist (e.g., "Buffer Overflow"). Ignore any extra/unrelated vulnerabilities listed by the detector. As long as the target is found, it is true.\\
- \texttt{INCORRECT:} The analysis fails to mention the target vulnerability type.\\

Dimension 2: Evidence (Insecure Code Snippet)\\
- \texttt{CORRECT:} The analysis quotes the exact code snippet (or a semantically identical variation) requested in the checklist.\\
- \texttt{PARTIALLY CORRECT:} The analysis cites the correct variable name or surrounding logic, but does not quote the specific snippet exactly. Or, it quotes a huge block of code that contains the snippet but lacks precision.\\
- \texttt{INCORRECT:} No specific evidence or incorrect code cited.\\

Dimension 3: Reasoning (Attack Mechanism)\\
- \texttt{CORRECT:} The analysis correctly explains both the root cause (e.g., "missing check") and the consequence (e.g., "overflow").\\
- \texttt{PARTIALLY CORRECT:} Explains the general issue (e.g., "unsafe copy") but misses technical details.\\
- \texttt{INCORRECT:} Incorrect or generic reasoning.\\

\textbf{3. Input}\\
The Ground Truth Checklist (Target: Vulnerable):\\
\{CHECKLIST\}\\

The Candidate Analysis:\\
\{ANALYSIS\}\\

\textbf{4. Output Format (Json Only)}\\
\verb|```|json\\
\{\\
\hspace*{1em}"Verdict\_Recall": \{\\
\hspace*{2em}"justification": "<Your brief reason>",\\
\hspace*{2em}"option": <choose from ["\texttt{CORRECT}", "\texttt{INCORRECT}"]>\\
\hspace*{1em}\},\\
\hspace*{1em}"Evidence\_Insecure\_Code": \{\\
\hspace*{2em}"justification": "<Your brief reason>",\\
\hspace*{2em}"option": <choose from ["\texttt{CORRECT}", "\texttt{PARTIALLY CORRECT}", "\texttt{INCORRECT}"]>\\
\hspace*{1em}\},\\
\hspace*{1em}"Reasoning\_Mechanism": \{\\
\hspace*{2em}"justification": "<Your brief reason>",\\
\hspace*{2em}"option": <choose from ["\texttt{CORRECT}", "\texttt{PARTIALLY CORRECT}", "\texttt{INCORRECT}"]>\\
\hspace*{1em}\}\\
\}\\
\verb|```|
\end{tcolorbox}
\caption{Specification-level Judge Prompt for Vulnerable Samples}
\label{box:sample_judge_vul}
\end{figure}

\begin{figure}[H]
\begin{tcolorbox}[boxrule=1pt, left=1mm, right=1mm, top=1mm, bottom=1mm, fontupper=\footnotesize]
\textbf{1. Goal}\\
Your task is to evaluate a vulnerability analysis for a code sample known to be \texttt{NO\_VUL} (vulnerability fixed). Verify if the detector correctly concludes the code is safe regarding the specific vulnerability.\\

\textbf{2. Scoring Rubric}\\
Dimension 1: Verdict (Strict Binary)\\
- \texttt{CORRECT:} The analysis says "\texttt{NO\_VUL}" or The analysis reports other vulnerabilities but does not list the specific fixed vulnerability mentioned in the checklist.\\
- \texttt{INCORRECT:} The analysis explicitly claims the specific target vulnerability (e.g., the one described in the checklist) still exists.\\

Dimension 2: Evidence (Safeguard Snippet)\\
- \texttt{CORRECT:} The analysis quotes the exact safeguard code (e.g., the new check/sanitizer) requested in the checklist.\\
- \texttt{PARTIALLY CORRECT:} The analysis cites the correct variable name or surrounding logic, but does not quote the specific snippet exactly. Or, it quotes a huge block of code that contains the snippet but lacks precision.\\
- \texttt{INCORRECT:} No specific evidence or incorrect code cited.\\

Dimension 3: Reasoning (Safety Logic)\\
- \texttt{CORRECT:} Explains why the code is safe (e.g., "The new check prevents the overflow").\\
- \texttt{PARTIALLY CORRECT:} Vague acknowledgment of safety without specific logic.\\
- \texttt{INCORRECT:} Incorrect logic or claims the code is unsafe.\\

\textbf{3. Input}\\
The Ground Truth Checklist (Target: Safe/Fixed):\\
\{CHECKLIST\}\\

The Candidate Analysis:\\
\{ANALYSIS\}\\

\textbf{4. Output Format (Json Only)}\\
\verb|```|json\\
\{\\
\hspace*{1em}"Verdict\_Absence\_of\_Specific\_Vuln": \{\\
\hspace*{2em}"justification": "<Your brief reason>",\\
\hspace*{2em}"option": <choose from ["\texttt{CORRECT}", "\texttt{INCORRECT}"]>\\
\hspace*{1em}\},\\
\hspace*{1em}"Evidence\_Safeguard\_Code": \{\\
\hspace*{2em}"justification": "<Your brief reason>",\\
\hspace*{2em}"option": <choose from ["\texttt{CORRECT}", "\texttt{PARTIALLY CORRECT}", "\texttt{INCORRECT}"]>\\
\hspace*{1em}\},\\
\hspace*{1em}"Reasoning\_Resolution": \{\\
\hspace*{2em}"justification": "<Your brief reason>",\\
\hspace*{2em}"option": <choose from ["\texttt{CORRECT}", "\texttt{PARTIALLY CORRECT}", "\texttt{INCORRECT}"]>\\
\hspace*{1em}\}\\
\}\\
\verb|```|
\end{tcolorbox}
\caption{Specification-level Judge Prompt for Patched Samples}
\label{box:sample_judge_fix}
\end{figure}
\end{document}